# Interface-Induced Stability of Nontrivial Topological Spin Textures: Unveiling Room-Temperature Hopfions and Skyrmions

F. Katmis[1,2,*,†], V. Lauter[3,*,†], R. Yagan[4], L. S. Brandt[5], A. M. Cheghabouri[4], H. Zhou[6], J. W. Freeland[6], C. I. L. de Araujo[5], M. E. Jamer[7], D. Heiman[2,8], M. C. Onbasli[4,9,†], & J. S. Moodera[1,2,†]

[1]Department of Physics, Massachusetts Institute of Technology, Cambridge, MA-02139, USA, [2]Francis Bitter Magnet Laboratory & Plasma Science and Fusion Center, Massachusetts Institute of Technology, Cambridge, MA-02139, USA, [3]Neutron Scattering Division, Neutron Sciences Directorate, Oak Ridge National Laboratory, Oak Ridge, TN-37831, USA, [4]Department of Electrical and Electronics Engineering, Koç University, Istanbul, 34450, Türkiye, [5]Departamento de Física, Universidade Federal de Viçosa, Viçosa, 36570-900, Brazil, [6]Advanced Photon Source, Argonne National Laboratory, Argonne, IL-60439, USA, [7]Physics Department, United States Naval Academy, Annapolis, MD 21402, USA, [8]Department of Physics, Northeastern University, Boston, MA 02115, USA, and [9]Department of Physics, Koç University, Istanbul, 34450, Türkiye

† To whom correspondence and request for materials should be addressed. Email: katmis@mit.edu (F.K.), lauterv@ornl.gov (V.L.), monbasli@ku.edu.tr (M.C.O.) and moodera@mit.edu (J.S.M.), *These authors contributed equally to this work.

**Abstract: Topological spin configurations, such as soliton-like spin texture and Dirac electron assemblies, have emerged in recent years in both fundamental science and technological applications. Achieving stable topological spin textures at room-temperature is crucial for enabling these structures as long-range information carriers. However, their creation and manipulation processes have encountered difficulties due to multi-step field training techniques and competitive interactions. Thus, a spontaneous ground state for multi-dimensional topological spin textures is desirable, as skyrmions form swirling, hedgehog-like spin structures in two dimensions, while hopfions emerge as their twisted three-dimensional counterparts. Here, we report the first observation of robust and reproducible topological spin textures of hopfions and skyrmions observed at room temperature and in zero magnetic field, which are stabilized by geometric confinement and protected by interfacial magnetism in a ferromagnet/topological insulator/ferromagnet trilayer heterostructure. These skyrmion-hopfion configurations are directly observed at room temperature with Lorenz transmission electron microscopy. Using micromagnetic modelling, the experimental observations of**

**hopfion-skyrmion assemblies are reproduced. Our model reveals a complete picture of how spontaneously organized skyrmion lattices encircled by hopfion rings are controlled by surface electrons, uniaxial anisotropy and Dzyaloshinskii-Moriya interaction, all at ambient temperature. Our study provides evidence that topological chiral spin textures can facilitate the development of magnetically defined information carriers. These stable structures hold promise for ultralow-power and high-density information processing, paving the way for the next generation of topologically defined devices.**

In order to minimize both size and power dissipation in future electronics, using spin degrees of freedom rather than strictly electronic charge, a durable and robust remanence will be required against external perturbation.[1,2] In addition, the desired stability to thermal fluctuations at room temperature may be achieved by topological protection of the states encoded in chiral spins. Thanks to their intriguing quantum features, topological insulators (TIs) hold immense promise for advancing technological applications to new heights.[3-17] When the ferromagnetic insulator (FMI) that surrounds the TI acquires a nonuniform magnetization, the interplay between Dirac electrons and the domain wall gives rise to a chiral state.[18-20] A unique interfacial chiral spin texture can be stabilized via coupling surface states with magnetic anisotropy thus bringing new functionalities to TI systems.[21-31]

Dissipation-free topologically protected surfaces of TIs serve as a unique platform for the formation of skyrmion-hopfion assemblies.[32-35] Similar to two-dimensional skyrmions, hopfions are soliton-like topologically protected spin textures that have been theoretically discovered and empirically confirmed in different material systems.[35-46] The origin of such localized helical magnetic textures is the competition between the Dzyaloshinskii-Moriya interaction (DMI) favouring non-collinearity and the Heisenberg exchange interaction favouring collinear alignments.[45-53] Achieving stable, topological multi-dimensional soliton-like spin textures and Dirac electron assemblies at room temperature remains challenging due to multi-step field training methods and competing interactions. To enable their use as long-range and dissipation-free information carriers at nanoscale, a spontaneous stable ground state for these textures is highly desirable.

We report the first observation of hopfion-skyrmion configuration at room temperature under zero applied magnetic field in a trilayer FMI-TI-FMI heterostructure of EuS-$Bi_2Se_3$-EuS thin film. Remarkably, these hopfion-skyrmion assemblies are observed far above the bulk Curie temperature of EuS, ~17 K. The chiral nature is driven by the interfacial DMI between adjacent FMI and TI, while the coupling strength at the two interfaces provides a tunable non-collinear spin texture as well as skyrmion lattice and hopfions. Through extensive characterizations and imaging, it is demonstrated that the chiral spin configuration exists at ambient temperature, even though the total magnetic moment reduces beyond the Curie temperature, evidently enabled by the strong DMI and interfacial coupling. Finally, the entire trilayer structure undergoes a phase transition between a ferromagnet and a unique combination of skyrmion and hopfion phase, which is confirmed by real space observation using a Lorentz transmission electron microscope (LTEM).

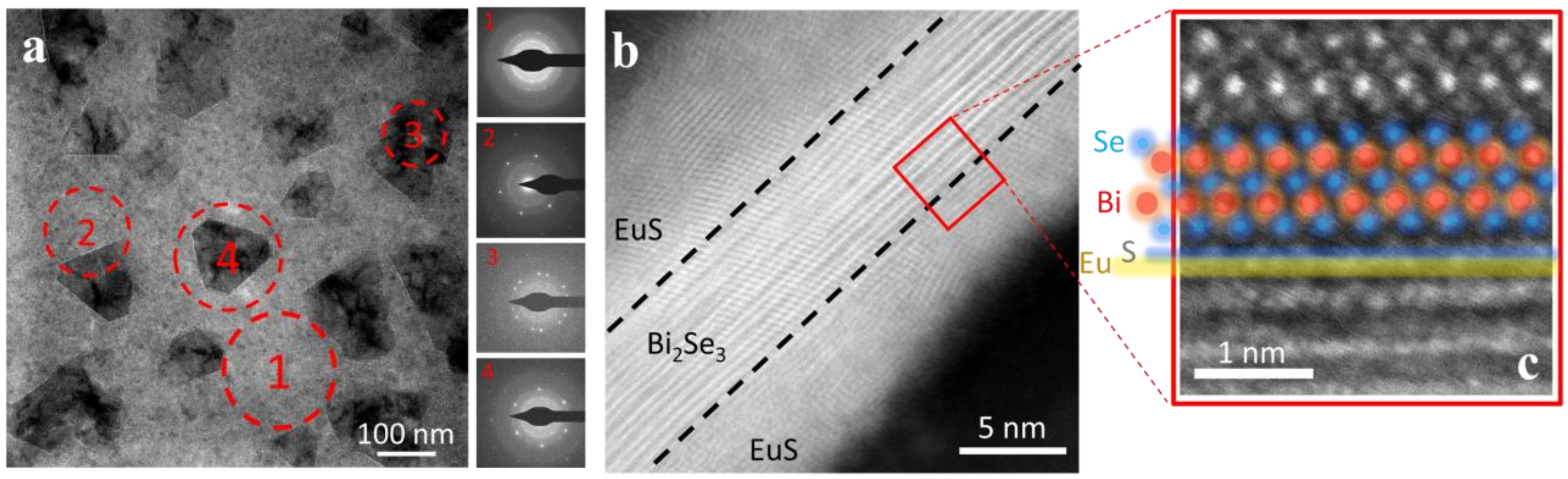

**Figure 1.** High-resolution TEM measurements for trilayer EuS-$Bi_2Se_3$-EuS (5 nm - 5 QL - 5 nm) grown on a $Si_3N_4$ membrane. In **a**, the HRTEM top-view image of a large area of a sample grown on a grid. The dark regions are single crystalline domains surrounded by polycrystalline lighter region, which are labelled as Region 1. A single domain island is labelled as Region 4. A closer look at a crystalline domain, where two domains merge to form a twin boundary labelled as Region 3. Selected area electron diffraction patterns of the corresponding regions (1, 2, 3, and 4) are shown between **a**, and **b**. In **b**, a cross-sectional bright-field TEM image of the trilayer sample on a membrane is shown, and an associated enlarged image of the highlighted region of one interface is indicative of well-oriented large crystallites in **c**.

Magnetically coupled EuS-$Bi_2Se_3$-EuS hybrid trilayer structures were grown using a well-optimized epitaxial technique.[8] The trilayer films were grown simultaneously on silicon nitride ($Si_3N_4$) membranes and on sapphire substrates to study the spatially- and depth-resolved magnetic configuration; further information is provided in the Materials and Methods section.

We performed high-resolution LTEM, which enables lateral magnetization mapping,[54,55] while the film quality was surveyed with conventional high-resolution TEM (HRTEM) techniques. The top view and cross-sectional view are shown in Figures 1a, and b, c, respectively; the spatial atomic distribution of a trilayer is shown in Figure S9 in the Supporting Information. The initial EuS layer deposited on the membrane exhibits a textured structure that serves as an effective seed layer for the subsequent growth of TI. This layer reduces the interface energy, promoting the two-dimensional growth of the TI layer. Over time, the TI layer evolves through the merging of smaller domains into larger domains, such coalescence resulting in either twin domains or a single expanded domain. To distinguish such domains, we labelled the regions; Region 1 is identified for polycrystalline phases and is confirmed by a broad ring electron diffraction pattern. Region 3 was taken on a twin domain. The cross-section cut shown in Figure 1b was made along a single-domain island as in Region 4. The trilayer configuration is clearly evident in the cross-section view. The area that has been highlighted reveals sharp interfaces between each atomic layer, as shown in Figure 1c. The position of the atoms, the van der Waals gap, and the EuS monolayers are all straightforwardly distinct. The interface between the EuS and $Bi_2Se_3$ layers is clearly defined, with an abrupt interface transition without chemical interdiffusion, according to the atomic density profile fitting. Typically, a weak van der Waals connection forms at the interfacial region between the S and Se layers located directly on top of each other with a typical gap size of $2.45 \pm 0.1$ Å, which is confirmed by an X-ray diffraction analysis for the layers grown on sapphire (See Figure S1 in the Supporting Information).

LTEM experiments were carried out on single crystalline domains of different sizes at room temperature, shown in Figure 2. In a zero magnetic field, we observe skyrmions arranged in a triangular-symmetric lattice with bright spots of 20±5 nm diameter at the vertices of the triangles and are spaced 35±5 nm apart from each other in Figure 2c. This type of spontaneous ordering is associated with the deflection of the electron beam due to the features of the curled magnetization of skyrmion core.[36,56] Observing ring like features, attributed to hopfion rings, surrounding the lattices of triangularly structured skyrmions is another significant finding.

It is remarkable that the dynamics and stability of skyrmions and hopfions may be influenced by one another. We observed that when an out-of-plane magnetic field is applied, the density of skyrmions increases, and hopfions become more organized inside each island, as shown in Figure 2a and b. From the figure, it is estimated that increasing the applied field decreases the skyrmion diameter down to ~10 nm before Heisenberg exchange-limited length-scale regime is reached, which is consistent with previously observed skyrmion sizes.[7,10] The value of the saturation field depends on the size of the island and can be estimated as 1 Tesla or more, which exceeds the LTEM bias capability. By examining different islands, it was found that the diameter of the skyrmions and the nature of the lattice formation were similar on different islands.

As can be seen in Figure 1, the trilayer creates crystalline domain islands, and between the islands there is a region with polycrystalline characteristics, which has a lower damping nature than the crystalline region. This magnetic and structural confinement opens the way for nucleation of hopfions at the boundaries of the islands. Owing to the strong magnetic properties of solitons, they were found to also act on the polycrystalline region, where standing wave patterns of spins are generated as a result of collective spin excitation.[57] As can be seen in Figure 2f and the Figure S2 in the Supporting Information showing how standing wave patterns of spins propagate with the magnetic field, the periodicity of the standing waves and the hopfion diameter decreases with increasing perpendicular field as in Fig. S2 b. Furthermore, as can be seen in Figure 2g with linear cuts in 2h along the three edges, we observe the interference pattern of standing wave patterns of spins excitations at a long-range scale in such independent islands (Figure S2, Supporting Information). The height profile analysis confirms that the observed magnetic phenomena are intrinsic to the EuS/$Bi_2Se_3$/EuS heterostructure, with line cuts along the red and black dashed lines in Figure S2 c-e revealing the characteristic truncated island morphology of epitaxially grown $Bi_2Se_3$ exhibiting thickness variations of approximately 2 QLs between adjacent structural units. LTEM demonstrates that both the magnetic contrast and the emergent hopfion-skyrmion assemblies maintain structural and magnetic coherence across islands despite these thickness fluctuations.

Our systematic investigation across various thickness combinations ($Bi_2Se_3$: 10-20 nm; EuS: 1-10 nm) demonstrates that interfacial spin textures remain qualitatively consistent, with the EuS thickness primarily affecting overall magnetic moment while maintaining constant interfacial exchange coupling strength per unit area. The magnetic proximity effect decays exponentially into $Bi_2Se_3$ with a characteristic length scale of ~1-2 nm, confining the induced spin texture to interface regions and allowing the opposite surface to approach intrinsic helical Dirac states beyond ~6-8 nm thickness. Laterally, $Bi_2Se_3$ forms truncated islands of 50-200 nm that serve as fundamental structural units, with magnetic domains (100-500 nm) creating scenarios where individual islands may contain single domains or span domain boundaries, resulting in intriguing multi-domain configurations. This island-limited geometry provides an advantageous framework for studying topological surface state interactions with magnetic proximity effects across both structural and magnetic boundaries while maintaining consistent behaviour within homogeneous regions.

To investigate the long-range magnetic order, continuous epitaxial trilayer EuS-$Bi_2Se_3$-EuS films were grown on a single crystal *c*-plane sapphire ($Al_2O_3$(0001)) substrate, see Materials and Methods for details. In epitaxial trilayers, bulk volume- and surface-sensitive techniques were used to confirm the existence of sharp chemical and electronic interfaces, as has been used in the case of bilayer samples.[8] The bottom and top EuS layers have six-fold symmetry due to the nature of the rotating domain, either on the substrates or on $Bi_2Se_3$, respectively. From the structural investigation as shown in Figure S1 in the Supporting Information, the bottom-up crystalline symmetry of the top EuS makes both layers structurally identical.

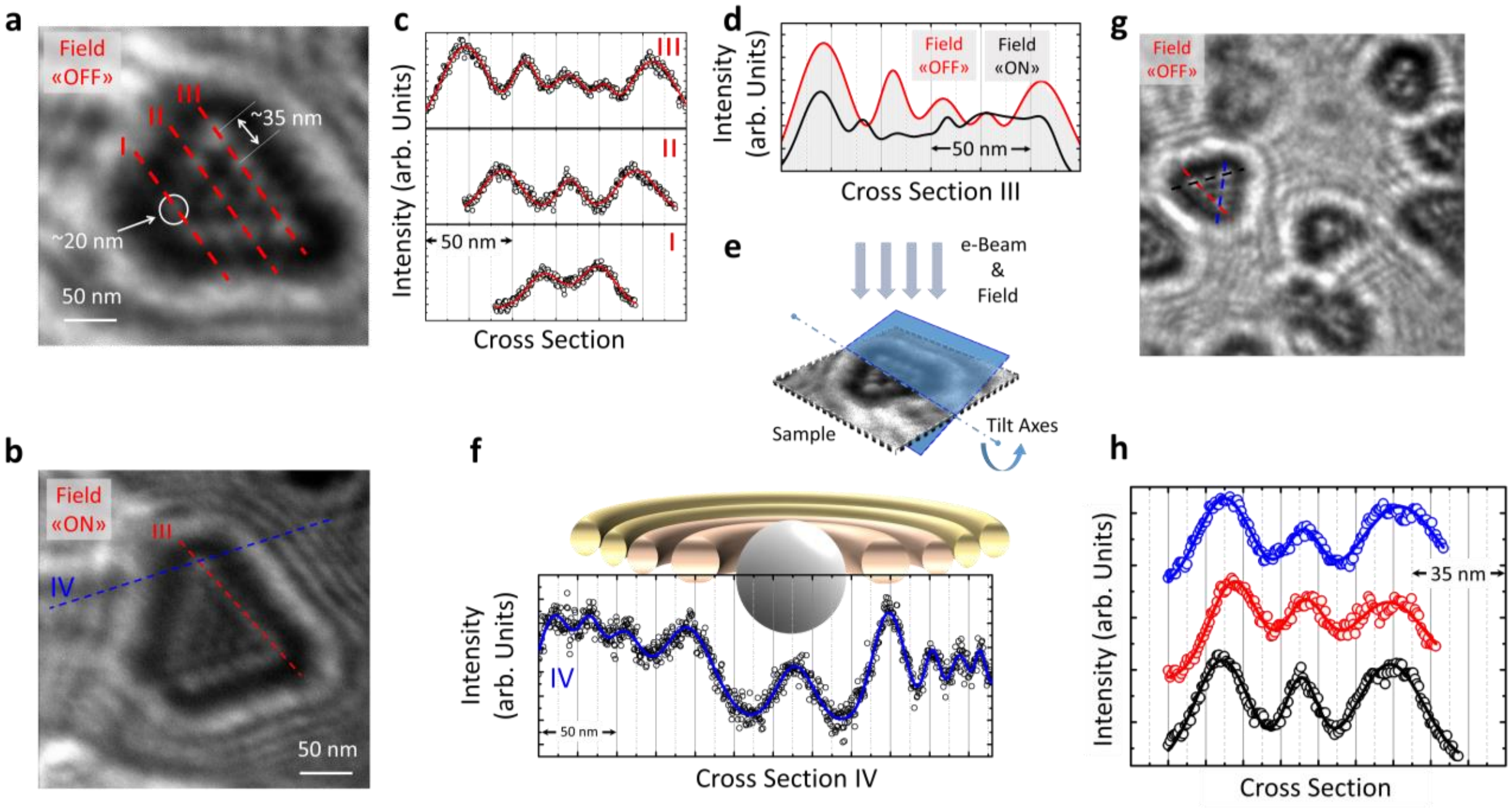

**Figure 2.** Lorentz TEM measurements at room temperature for trilayer EuS–$Bi_2Se_3$–EuS (5 nm – 5 QL – 5 nm) grown on $Si_3N_4$ membrane. The LTEM was performed when the field is ON in **a,** and OFF in **b** at room temperature. A closer look at one of the truncated islands, where the triangular skyrmion lattice is formed with a periodicity of 35 ± 5 nm with the size of each skyrmion 20 ± 5 nm. In **c**, the distinct skyrmion lattice line cut is seen along I, II, and III in **a**. In **d**, the magnetic field applied out-of-plane by the microscope objective lens when the sample is tilted shows an increase in the density of skyrmion in the lattice as in the histogram data. The applied out-of-plane component of the magnetic field is 1.53 T, "Field ON" condition. The periodicity is shown as the integrated intensity histogram of the lateral cross-sectional along III without field (in red) and with field (in black). In **e**, a schematic illustration of the imaging configuration while manipulating the sample under a magnetic field. In **f**, standing wave patterns of spins outside of the crystal region are plotted along the line intersecting IV in **b**. These peaks demonstrate the presence of skyrmion lattice points at the edge, along with the hopfion ring. Additionally, clear spin wave formations surrounding the island exhibit wave-like characteristics that are clearly visible in **b**. Above it is the schematic graphic that serves as a guide for the eyes. In **g**, a closer look at one of the smaller size triangular islands where the perfect triangular lattice is formed with the same periodicity. The integrated intensity histogram for the island seen in **g** is displayed along each of the three side edges in **h**.

Although proximity-induced magnetic phenomena are apparent from the macroscopic magnetization measurements (see for example, Figures 3a and b), the contribution of interfacial magnetism between EuS and $Bi_2Se_3$ at low temperatures is small compared to the bulk magnetism of EuS. Interfacial effects at high temperatures become more apparent when bulk magnetism mostly vanishes above the Curie temperature.[8]

Element-specific X-ray absortion (XAS) and X-ray magnetic circular dichroism (XMCD) measurements were performed using total electron yield (TEY) detection mode, which provides enhanced surface/interface sensitivity compared to bulk fluorescence yield. XMCD spectra for the $Eu^{2+}$ state as a function of temperature show a sign of significant interfacial coupling, as presented in Figure S10 in the Supporting Information. While XMCD at Se L-edges showed magnetic dichroism, spectral overlap with one of the Eu edges prevents unambiguous attribution to Se magnetization. The observed interfacial magnetic proximity effects are confirmed through complementary polarized neutron reflectometry analysis. It is crucial to emphasize that due to the highly localized nature of the exchange interaction, achieving an exceptionally sharp, well-defined, and clean interface between EuS and TI is imperative to ensure the necessary effective magnetic proximity coupling. Therefore, careful handling is essential during their formation and subsequent magnetic interfacial structuring in order to minimize any unwanted contamination from leftover chalcogen atoms.[58]

For elucidating the spin texture and the special distribution of magnetism in the trilayer films, we employed the depth-sensitive polarized neutron reflectometry (PNR) to reveal the non-collinear magnetic order along the two interacting heterointerfaces. In the case of bilayer samples, the ferromagnetism extends to about 2 nm into the TI layer, while due to such short-range nature the time-reversal symmetry could only be broken in the vicinity of the interfacial region.[8] When an additional magnetic interface is created near to the other interface, close enough to interact, the total magnetic configuration changes dramatically. To overcome the anisotropy, the PNR reflectivity profiles $R^+$ and $R^-$ were measured at 300 K in saturating in-plane external magnetic field of 1 T and 5 T. PNR results reveal that the two coupled interfaces make both EuS layers in FMI-TI-FMI trilayer totally magnetic to much higher temperatures, the corresponding depth profiles of neutron structural, magnetic and absorption scattering length density (NSLD as green, MSLD as grey, and ASLD as purple)  are shown in Figure 3e for 300 K, and 5 K in Figure S3 in the Supporting Information. The corresponding spin asymmetry (SA) data are shown in Fig. 3c and 3d for 1 T and 5 T, respectively. As an indicator of Eu atoms, the ASLD depth profile[59] ends right at the $Bi_2Se_3$ interface, indicating that no Eu atoms were found in the $Bi_2Se_3$ layer. Consequently, PNR provides concrete proof that these trilayer heterostructure exhibits room-temperature ferromagnetism produced by proximity. Depth-sensitive PNR reveals that in tri-layer EuS/$Bi_2Se_3$/EuS heterostructures, both EuS layers maintain magnetization throughout their entire 5 nm thickness up to 300 K, representing a significant enhancement over bi-layer samples where magnetization was retained only in a narrow ~2 monolayer of EuS interfacial region. This suggests that the sandwiched $Bi_2Se_3$ layer acts as a magnetic mediator, extending interface-driven magnetic stabilization through long-range coupling mechanisms, with the double-interface geometry providing stronger magnetic coupling than conventional bi-layer configurations.

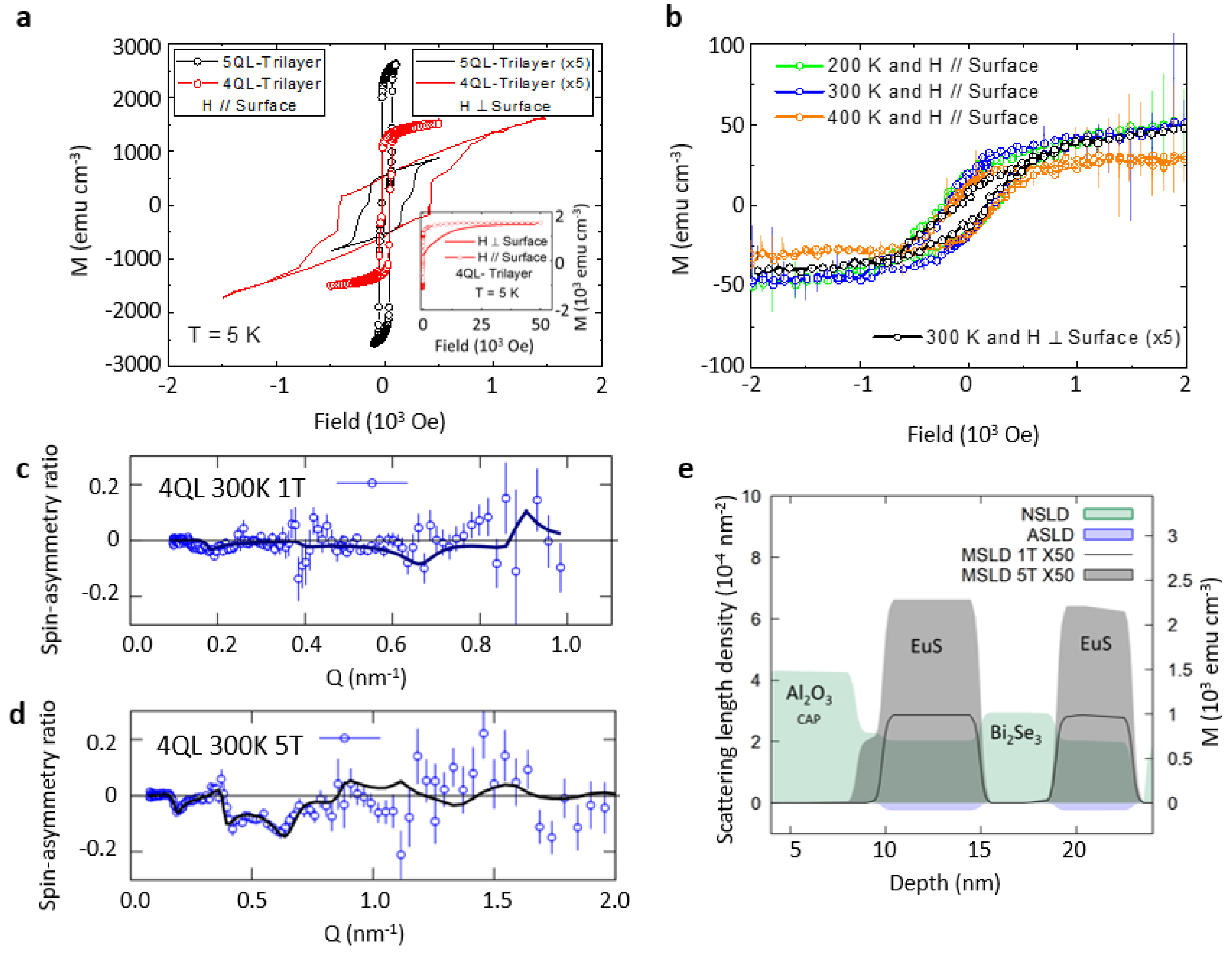

**Figure 3.** SQUID magnetometry and PNR measurements for trilayer EuS–$Bi_2Se_3$–EuS, In **a**, measurements of **M**(**H**) at low field and temperature in a parallel and perpendicular field configuration for 4 QL-trilayer (EuS (5 nm) – $Bi_2Se_3$ (4 QL) – EuS (5 nm)) and 5 QL-trilayer (EuS (5 nm) – $Bi_2Se_3$ (5 QL) – EuS (5 nm)) samples. The **M**(**H**) at high field region for a 4-QL trilayer sample with parallel and perpendicular arrangement is displayed in the inset. **b** shows the **M**(**H**) for the 5 QL trilayer at high temperatures in parallel and perpendicular field configurations. The corresponding spin-asymmetry (SA) ratio and model fits shown by solid lines, $SA = (R^{+}-R^{-})/(R^{+}+R^{-})$, derived from the simultaneous fitting of the polarized neutron reflectivity measurement shown for 4 QL-trilayer samples under 1 T in **c** and 5 T in-plane field in **d**. **e** displays neutron nuclear (NSLD, green), magnetic (MSLD, grey) and absorption (ASLD, purple) scattering length density profiles across the trilayer epitaxial sample which were recorded at 300 K with in-plane field of 1 and 5 T.

Using micromagnetic models based on the Landau-Lifshitz-Gilbert (LLG) formalism and aided by the LTEM, SQUID and PNR data, the origin of protected chiral magnetism at various temperatures has been examined for the occurrence of the magnetic texture skyrmion-hopfion assembly (See Methods section for details in the supplementary file). Experimental observations of the chiral spins were closely reproduced using micromagnetic modelling and their numerical solutions. Phase diagrams of hopfion with skyrmion lattice under different geometric confinement have been generated in a broad parameter space, such as different DMI settings, uniaxial anisotropy parameters, and saturation magnetizations.

The effects of changes in the DMI and saturation magnetization $M_{sat}$ on the stability of a single skyrmion island are illustrated in Figure S4 in the Supporting Information, where the initial settings without hopfion are displayed. The results can be categorized into three areas: (1) A skyrmion with an out-of-plane magnetic orientation for lower $M_{sat}$ and lower DMI can occur in certain areas. (2) A skyrmion might still be stabilized under different magnetic parameter combinations and increased $M_{sat}$ for the in-plane film magnetization. (3) The increased DMI dominates the system and produces a labyrinth-like magnetism because of the decreased anisotropy and exchange energy. Note that the parameter window $M_{sat}$ and DMI that stabilizes the hopfion with skyrmion lattice in the models are consistent with the earlier experimental results.[8]

Figure 4 illustrates how the formation of skyrmion and hopfion rings are influenced by the geometrical confinement, with island sizes of Figure 4a 100, Figure 4b 300 and Figure 4c 500 nm$^2$. The geometrical size defines the scale of the hopfion and skyrmion lattice. This imposes strong demagnetization fields, and the symmetry is broken at very small geometric sizes, as in case of the 100 nm island, in Figure 4a, due to competition between the edges and magnetic features. Furthermore, after relaxation of the model, the influence of the *DMI* constant and the saturation magnetization on the stability of the hopfion ring protecting the skyrmion lattice is demonstrated. The ring diffuses into the triangle geometry's borders at low *DMI* and $M_{sat}$, and the magnetic features' topology is lost. The hopfion ring remains unbroken at higher *DMI* and $M_{sat}$, as particularly seen at $DMI = 4.56\times10^{-6}$ J m$^{-2}$ and $M_{sat} = 37.6$ kA m$^{-1}$. This is also demonstrated for the reduced number of lattice-forming skyrmions, shown in the Figure S6 in the Supporting Information, similar to the observed features in the experimental data of Figure 2g. By rigorously checking the derived parameters during both the coarse (Figure S7, Supporting Information) and fine (Figure S8, Supporting Information) evolution phases, a comprehensive analysis of the entire parameter space was performed. This study revealed a relatively narrow operating regime that favours the formation of skyrmion-hopfion assemblies.

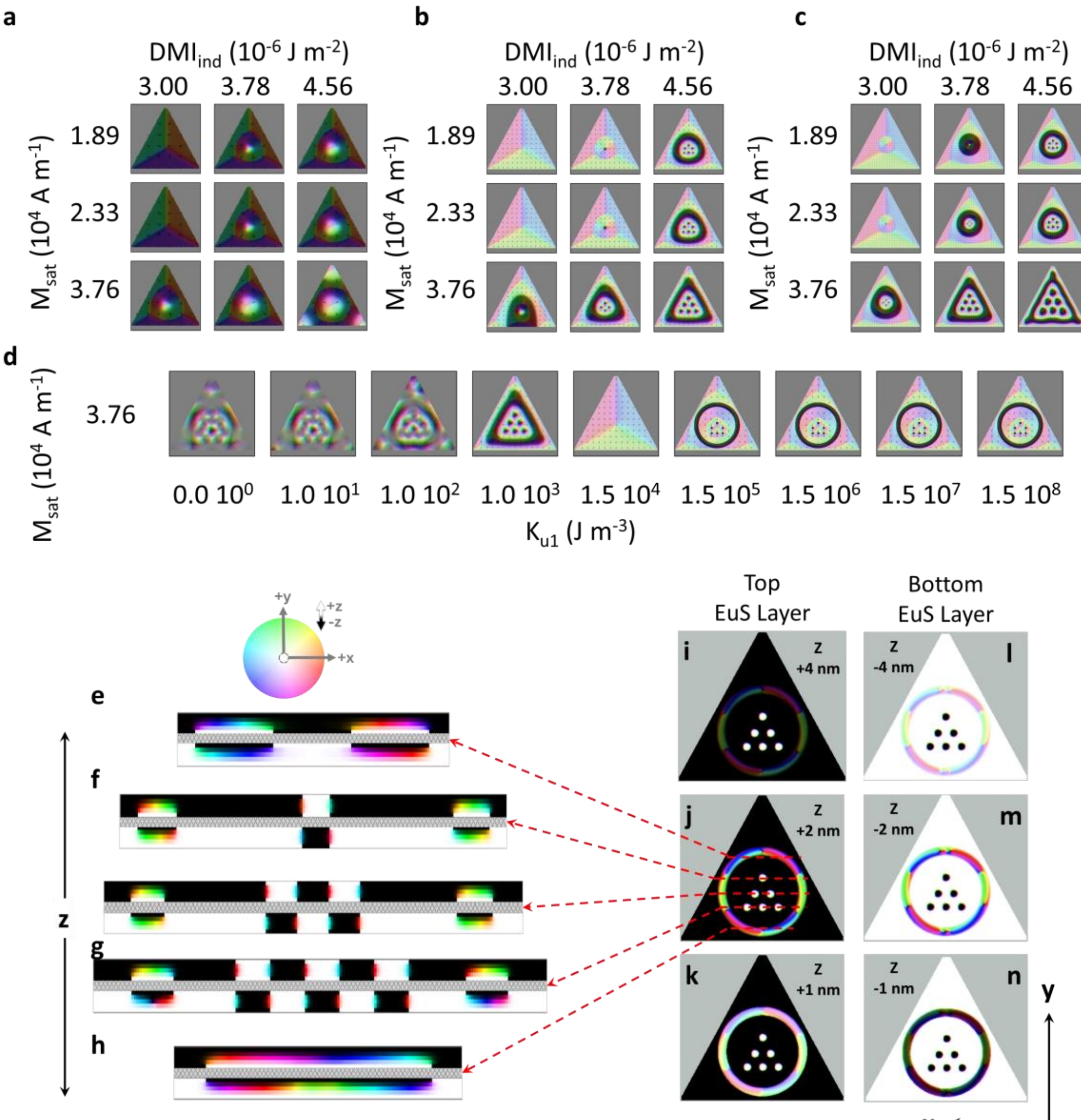

**Figure 4.** Micromagnetic simulations of magnetic texture with hopfion rings for trilayer films. The corresponding simulations for Figure 2g were run on regular triangle geometry with sizes at 100 in **a**, 300 in **b**, and 500 nm$^2$ in **c**. The entire isosurfaces of the skyrmion and hopfion lattice construction are matched with experimental data in **b**, due to their accurate size and shape by convenient $M_{sat}$ and DMI parameters. Skyrmion lattice surrounded by a hopfion ring initialized in a uniform domain of +**m** (white) in a triangular geometry as features of -**m** (black) and ran a simulation to obtain a final relaxed output. In **d**, a large range was also run for the corresponding uniaxial anisotropy parameters $K_{u1}$ with $DMI_{ind} = 4.56 \times 10^{-6}$ J m$^{-2}$, $A_{ex} = 1.94 \times 10^{-14}$ J m$^{-1}$, and $M_{sat} = 37.6$ kA m$^{-1}$ with regular triangle geometry with sizes at 300 nm$^2$. The phase profile along the x-z plane at different y positions is displayed in **e** to **h**, while the lateral view is depicted

in the x-y plane is shown in **i** to **n**. Lateral cross-sections were extracted at different EuS thicknesses, specifically at ±1, ±2, and ±4 nm, corresponding to their distances from each $Bi_2Se_3$ interface, where the + and - signs indicate positions within the top and bottom EuS layers, respectively. The color bar represents the spin phase changes corresponding to increasing azimuthal and polar angles.

From our micromagnetic analysis, the ring in-plane diameter is ~ 21 nm, and each skyrmion has a diameter of 20-25 nm for various numbers of confined skyrmions and independent of geometries as in the experimental observation. These results suggest that the fingerprint of the topological magnetic ordering is unique for the geometric confinement and intrinsic magnetic parameters. For zero or low uniaxial anisotropy constant $K_{u1}$ (0-$10^2$ J $m^{-3}$), the magnetic features exhibit a diffused ground state as shown in Figure 4d. When $K_{u1}$ is increased to $10^3$ J $m^{-3}$ in the out-of-plane direction (+**z**), the magnetic topological order exhibits the skyrmion-hopfion assembly that resembles the LTEM results shown in Figure 2g. In the high $K_{u1}$ regime ($10^5$-$10^8$ J $m^{-3}$), the uniaxial anisotropy energy term starts dominating over the other magnetization terms including *DMI*, demagnetization or geometry effects, intralayer exchange ($A_{ex}$), interlayer exchange and Ruderman-Kittel-Kasuya-Yoshida (RKKY) interaction. The hopfion and the skyrmion lattice could be stabilized in the micromagnetic models when the exchange, in-plane demagnetization, and the *DMI* fields are selected.

Previous studies suggested a similar stabilization mechanism in which RKKY,[60] the Bloembergen-Rowland interaction,[61] spin-orbit coupling,[62] and antiferromagnetic coupling on the surface help to stabilize the interface magnetism.[63] Micromagnetic models show that antiferromagnetic chiral ordering is stabilized for both interfaces at the same time when the RKKY energy term is included. The fact that an RKKY term is needed for the stability implies that the magnetism at both EuS-$Bi_2Se_3$ interfaces is likely coupled. The phase diagrams provide evidence that the stable magnetic topological ordered features could be achieved under the explored material parameter space, as shown in Figure S11 in the Supporting Information.

Our micromagnetic simulations predict Néel-type skyrmions in the ultra-thin film limits, and we have ensured that the simulation closely match those of the experimental conditions.[51,55,64] The Fresnel oscillations that are observed consistently in both our simulations and experiments align with the expected features for Néel-type skyrmions, especially in the presence of interfacial DMI. We cannot observe bulk/Bloch skyrmions to be stabilized in ultra-thin film limit as in phase profile analysis Figure 4e-n. Once the thickness increases to 20 nm EuS, we show that bulk/Bloch skyrmions are obtained in our micromagnetic models as shown in Figure S12 in the Supporting Information. The skyrmion phase profile observed in Figure S12 in the Supporting Information is consistent with the definition of the bulk/Bloch skyrmion.[65]

Our analysis reveals topological features consistent with hopfions, clearly distinguishing them from skyrmion bags. Specifically, our system consists of ultra-thin layers, only a few nanometers thick, where DMI arises exclusively from interfacial effects. The dimensional confinement in our system gives rise to unique magnetic configurations that differ significantly from those observed in bulk-like materials.[66] The simulated and analyzed structures are governed by this interfacial DMI and strong in-plane demagnetizing fields, resulting in the confinement of the spin phase profile of hopfions to within a few unit cells. To clarify, we have performed a detailed three-dimensional analysis of the magnetic textures in our simulations, and a thorough investigation of the magnetization profiles along all spatial dimensions as shown in Fig. 4 and S12. These analyses clearly demonstrate that the observed structures possess a non-trivial topology, which is characteristic of hopfions, rather than mere stacking of skyrmion bags.

A systematic comparison of EuS/$Bi_2Se_3$/EuS heterostructures and bulk-like FeGe systems reveals significant differences in the demagnetizing field effects due to the ultrathin nature of the films. In contrast to bulk FeGe,[33] these nanometer-scale structures exhibit significantly modified demagnetizing fields. Analysis of radial, azimuthal, and polar spin phase profiles in heterostructures with 20 nm and 5 nm thick EuS layers confirms the presence of fiber-like nontrivial topological configurations, including hopfion rings and skyrmions. Increasing the film thickness reduces the strong in-plane demagnetizing fields, facilitating appearance of swirling or fiber-like topological textures of hopfions, distinctly different from those in bulk FeGe. The dimensional confinement in these thin films induces unique magnetic configurations, governed by the interfacial DMI and strong in-plane demagnetizing fields, which confine the spin phase profile of the hopfions to a few unit cells. This extreme spatial confinement challenges the visual reconstruction of hopfion geometries compared to the more flexible formation in bulk FeGe, highlighting the critical influence of interfacial and confinement effects on topological spin textures.

The interface coupling also depends on the initial configuration, whereas the magnetic topological ordered structures are stable once initialized in the forms indicated in the phase diagrams. Despite the large in-plane demagnetizing field in the nanoscale geometric confinement of the irregular shape trilayers in the remanent state, the robust stability of the hopfion with skyrmion lattice at room temperature reported in other studies[67,68] suggests that these additional energy terms will be necessary to maintain the equilibrium lattice at room temperature. It has been discovered that, for the parameter choices considered in the models, the topological order is unstable at room temperature. Reliable prediction of magnetic order at room temperature is hampered by the uncertainty in magnetic parameters such as DMI, RKKY, and other uniaxial or strain-related anisotropy constants, as well as the crucial role of surface states. Nevertheless, the room temperature magnetic order can be partially reproduced when the uniaxial anisotropy parameter is on the order of $10^7$ and $10^8$ J $m^{-3}$.

The geometric confinement identified in our study enables room-temperature zero-field stability by providing a controlled platform to demonstrate the underlying interfacial phenomena. This confinement reveals the fundamental stabilization mechanisms governing topological spin textures at ferromagnet-TI interfaces, where long-range magnetic ordering extends throughout the entire ferromagnetic layer thickness. Our findings illuminate the fundamental physics governing topological structure stabilization, providing pathways to engineer wider area interfacial systems through alternative approaches such as interfacial engineering or strain modulation that replicate the beneficial effects of geometric confinement.

The first-principles study[63] identified magnetic proximity effects in trilayer EuS/$Bi_2Se_3$/EuS systems, demonstrating exponential decay of interlayer exchange with topological insulator thickness. While the concept of hidden DMI suggests opposite-sign interfacial DMI at top and bottom interfaces due to inversion symmetry,[69,70] experimental validation shows DMI is intrinsically short-range and localized within 1–2 nm of FMI/TI interfaces.[11,71] Given our ~5 nm $Bi_2Se_3$ spacer thickness, the interfacial DMI fields at opposite surfaces remain effectively independent without direct coupling, while interface-level asymmetries from domain boundaries and crystallographic imperfections naturally break the inversion symmetry conditions required for hidden DMI cancellation.

These findings demonstrate the capability to generate multi-dimensional spin textures, such as hopfions and skyrmions, on topological insulator surfaces through precise control of interfacial coupling and geometric confinement. The combination of experimental observations and micromagnetic modelling reveals that interfacial Dzyaloshinskii-Moriya interactions and spin-momentum locking at the topological insulator interface are key to stabilizing these textures at room temperature. These spin structures potentially enable low-dissipation information manipulation, which could, with further engineering advances in electrode design and operating conditions, eventually contribute to more energy-efficient data storage and transmission technologies. The ability to engineer stable spin textures through interface design offers a pathway to integrate these topologically protected states into spintronic devices, where they could enable low-power, high-speed operations. Moreover, the potential interaction of these textures with quantum states, such as Majorana fermions, opens avenues for topological quantum computing. The robustness of these structures at ambient temperatures positions them as promising candidates for practical applications in next-generation memory, logic, and sensing devices.

**Acknowledgements**

F.K. thanks L. Fu, V. Madhavan, P. Böni, K. Moler, P. Jarillo-Herrero, N. Gedik, and Y. Wang for discussions. This work was supported by the Army Research Office (ARO W911NF-20-2-0061), the National Science Foundation (NSF-DMR 2218550), Office of Naval Research (N00014-20-1-2306). F.K. and J.S.M. thank the Center for Integrated Quantum Materials (NSF-DMR 1231319) for financial support. This work made use of the MIT Material Research Laboratory. V.L. thanks H. Ambaye for partial assistance during the experiment. This research used resources at the Spallation Neutron Source, a Department of Energy Office of Science User Facility operated by the Oak Ridge National Laboratory. D.H. thanks the National Science Foundation for support by the National Science Foundation grant DMR-1905662 and the Air Force Office of Scientific Research award FA9550-20-1-0247. The use of the Advanced Photon Source was supported by the US Department of Energy, Office of Science, Office of Basic Energy Sciences, under contract number DE-AC02-06CH11357. M.C.O., R.Y. and A.M.C. acknowledge the European Research Council (ERC) Starting Grant SKYNOLIMIT with No. 948063 and ERC Proof of Concept project SuperPHOTON with No. 101100718. L.S.B. thanks INL-Braga in Portugal for providing the LTEM setup. C.I.L.A. acknowledges Brazilian agencies FINEP, FAPEMIG APQ-04548-22, CNPq, and CAPES (Finance Code 001).

**Conflict of Interest**

The authors declare no conflict of interest. Correspondence and requests for materials should be addressed to F.K. (katmis@mit.edu) and J.S.M. (moodera@mit.edu).

**Author Contributions**

The research was conceived and designed by F.K. and J.S.M. The samples were prepared and characterized by F.K. The high-resolution TEM and Lorentz TEM experiments were carried out by L.S.B and C.I.L.A. and analysed by F.K.; the PNR experiments and data analysis were carried out by V.L.; the XAS/XMCD experiments and data analysis were carried out by F.K. and J.W.F.; the XRD experiments and data analysis were carried out by F.K. and H.Z.; the SQUID experiments and data analysis were carried out by F.K., M.E.J., and D.H.; and R.Y., A.M.C., and M.C.O. developed the micromagnetic model, performed numerical simulations, and analysis done by R.Y., A.M.C., M.C.O., and F.K. The data was interpreted by F.K., V.L., M.C.O. and J.S.M. All authors discussed the results and commented on the manuscript. The manuscript was written by F.K, V.L. and M.C.O. with contribution from all authors.

## MATERIALS AND METHODS

**Material growth:** The epitaxial growth of both trilayers (EuS–$Bi_2Se_3$–EuS) grown on sapphire and membrane were performed simultaneously. The entire process was carried out in a custom-built molecular beam epitaxy (MBE) setup equipped with an electron beam evaporator for EuS deposition and effusion cells with 5N (99.999%) purity Bi and Se at a base pressure of $2 \times 10^{-10}$ Torr. As a final stage, all multilayers were protected in situ with amorphous $Al_2O_3$ layers. From the bottom up, each interface formation and structural evolution were monitored by an in situ reflection high-energy electron diffraction (RHEED) apparatus. Due to the high reactivity of Eu atoms and dissociation problems of S, the initial EuS was evaporated congruently from a single electron-beam source at 260 ± 5 °C with a rate of 1.0 - 1.5 Å $s^{-1}$. After the first EuS layer growth, the temperature of the effusion cells was increased to prevent Se incorporation into the layer. The temperature of the substrate was maintained at 240 ± 5 °C, while the temperature of the cells was raised; this procedure takes approximately one hour. Bi and Se were then co-evaporated with a flux ratio of 1:15 A growth rate of 1 - 2 Å $min^{-1}$ was utilized to create an ultra-smooth surface in order to prevent kinetic surface roughening. The temperature of the effusion cells was lowered immediately after $Bi_2Se_3$ growth. The growth stage was maintained at growth temperature during the cooling down process of the cells. As the temperature of the cells dropped, the second EuS layer was deposited with the same growth methodology as the initial layer. Finally, in the same deposition chamber, 5-10 nm amorphous $Al_2O_3$ was deposited in situ as a protective layer to all trilayer films. An energy dispersive spectroscopy (EDS) map showing the trilayer sample grown on the $Si_3N_4$ membrane's spatial atomic distribution. Based on the element mapping as shown in Extended Data Fig. 9, the elements Bi, Se, Eu, and S are uniformly distributed across the film surface.

**Structural Analysis:** Crystal-truncation rod (CTR) measurements analysed by coherent Bragg rod analysis (COBRA) method was used to render the real space electron density profile across the interfacial area with atomic accuracy, in an effort to shed light on the interface microstructure[72]. The CTR measurements were carried out at the Beamline 12-ID-D of Advanced Photon Source (APS), Argonne, using a six-circle diffractometer with an X-ray energy of 20 keV (wavelength $\lambda$ = 0.6199 Å) beam. Using the COBRA approach, the sample's specular CTR (00L) for both the bilayer and trilayer film configurations was examined. The interface between EuS and $Bi_2Se_3$ layers, as well as between $Bi_2Se_3$ and sapphire single crystal, are clearly delineated from the total electron density profile. Resolving the interface bond clearly demonstrates a sharp interface transition in the absence of any chemical inter-diffusion. Se atoms are in direct contact with either the S layer on EuS or the last oxygen layer of sapphire for bilayer films. $Bi_2Se_3$ is terminated at each interface by the Se layer, and the S layer is directly stacked on top of the Se layer by interfacial van der Waals bonding spacing of 2.45 Å. Overall, an excellent epitaxial cube-on-hexagon growth connection is seen for EuS (111) on $Bi_2Se_3$ and sapphire.

**SQUID magnetometry measurements:** were taken to observe the magnetization state of the multilayer systems and performed in a Quantum Design MPMS superconducting quantum interference device (SQUID) magnetometer. Both in-plane and out-of-plane magnetic properties were measured in the temperature range of 5–400 K and applied magnetic fields up to 5 T.

**Lorentz Microscopy measurements**[73]: were performed on samples grown on silicon nitride grids at room temperature. The good quality of trilayer structure was achieved in well-defined grains. From the TEM images it is possible to note regular arrays of helical magnetic structures with lattice parameter depending on the grain geometry. The Cs-probe corrected transmission electron microscope FEI Titan Themis 80–300 kV, was employed to perform high-resolution transmission electron microscopy (HRTEM), energy dispersive spectroscopy (EDS), and Lorentz transmission electron microscopy (LTEM) using Fresnel method (out-of-focus). The experiments were carried out at 300 kV and RT. A special Lorentz lens allows LTEM imaging in magnetic field-free conditions when the standard objective lens is switched off. By appropriately increasing the objective lens excitation, a magnetic field arises and is perpendicular to the sample surface. However, before exciting the objective lens, in order to apply a magnetic field to the sample, it was tilted up to 40°. For the "Field-ON" configuration shown in Fig. 2, the applied out-of-plane component of the magnetic field is 1.53 T. Specifically, the defocus values were 34 µm for conventional TEM imaging and from 0.89 mm to 1.38 mm for magnetic imaging.

Exit wave function of diffracted electrons in LTEM (non-magnetic crystal, no external magnetic field): To capture the Fresnel oscillations of transmitted electrons in LTEM, we initially define the exit wave function of electrons, considering the periodicity of the nonmagnetic crystal structure. Subsequently, we incorporate the spin structure, which exhibits an incommensurate periodicity relative to the reciprocal wave vector of the lattice.

The translational symmetry of the crystal brings a reciprocal lattice wave vector **G**. In a crystal with a periodic potential, $V(\mathbf{r}) = V(\mathbf{r}+\mathbf{a})$, where **a** is a lattice vector. This periodic lattice potential leads to the Bloch wave solutions with wave vector **k** and $u_{\mathrm{k}}(\mathbf{r}) = u_{\mathrm{k}}(\mathbf{r}+\mathbf{a})$; $\psi(\mathbf{r}) = u_{\mathrm{k}}(\mathbf{r})e^{i\mathbf{k}\cdot\mathbf{r}}$. The exit wave function in LTEM becomes: $\psi_{\mathrm{exit}}(\mathbf{r}) = \psi_{\mathrm{incident}}(\mathbf{r})e^{i\phi(\mathbf{r})}$, where $\phi(\mathbf{r})$ is the phase shift introduced by the crystal potential. For a thin crystal of thickness t, $\phi(\mathbf{r})$ is

$$\phi(\mathbf{r}) = \frac{\sigma}{\hbar v}\int_0^t V(\mathbf{r}, z)\, dz$$

Fresnel oscillations arise due to the interference of multiple scattered waves. A crystal potential with a Fourier expansion in the reciprocal lattice can be used to describe the transmitted electron wave function. Thus, the transmitted wave can be written as a sum of diffracted waves:

$$\psi_{\text{trans}}(\mathbf{r}) = \sum_{\mathbf{G}} A_{\mathbf{G}} e^{i(\mathbf{k}+\mathbf{G})\cdot\mathbf{r}}; \qquad I(\mathbf{r}) = |\psi_{\text{trans}}(\mathbf{r})|^2 = \left|\sum_{\mathbf{G}} A_{\mathbf{G}} e^{i(\mathbf{k}+\mathbf{G})\cdot\mathbf{r}}\right|^2$$

The amplitudes of the diffracted waves $A_{\mathbf{G}}$ are determined by the scattering conditions, which depend on both sample and measurement parameters. Here, $I(\mathbf{r})$ is the intensity observed in LTEM. This expression contains the interference terms $A_{\mathbf{G}}A^*_{\mathbf{G}'}$ and $e^{i[(\mathbf{G}-\mathbf{G}')\cdot\mathbf{r}]}$, which lead to the oscillatory patterns in the intensity corresponding to the LTEM Fresnel oscillations of nonmagnetic crystals.

When we introduce the spin order with wave vector $\mathbf{q}$ that is incommensurate with the crystal wave vector $\mathbf{G}$ and $V_{\mathbf{q}}$ are the Fourier components of the spin-modulated magnetic potential: $V_{\text{spin}}(\mathbf{r}) = \sum_{\mathbf{q}} V_{\mathbf{q}} e^{i\mathbf{q}\cdot\mathbf{r}}$. For a magnetic crystal with spin ordering, the total potential contains the contributions from the periodic lattice potential $V_{\text{lattice}}(\mathbf{r})$ and the spin wave potential $V_{\text{spin}}(\mathbf{r})$; $V_{\text{total}}(\mathbf{r}) = V_{\text{lattice}}(\mathbf{r}) + V_{\text{spin}}(\mathbf{r})$. When we apply external magnetic field, the Schrödinger equation must be modified to include the Aharonov-Bohm phase, which includes the vector magnetic potential ($\nabla \times \mathbf{A} = \mathbf{B}$) that modifies the electron wavefunction's phase. The Zeeman interaction term also modifies the spin-dependent potential:

$$-\frac{\hbar^2}{2m}\left(\nabla - i\frac{e}{\hbar}\mathbf{A}\right)^2 \psi(\mathbf{r}) + V_{\text{total}}(\mathbf{r})\psi(\mathbf{r}) = E\psi(\mathbf{r})$$

Under a constant magnetic field $\mathbf{B} = B\hat{z}$, the vector potential can be expressed in the symmetric gauge as $\mathbf{A} = ½\mathbf{B} \times \mathbf{r} = ½\, B\,(-y\hat{x} + x\hat{y})$. The Zeeman interaction accounts for the coupling of electron's magnetic moment to external magnetic field. This interaction modifies the different spin wave potential from $V_{\text{spin}}(\mathbf{r})$ to $V_{\text{spin}}(\mathbf{r}) + \Delta V_{\text{Zeeman}}$. The modified wave function is now subject to the total potential from the lattice, spin wave and the Zeeman contribution: $V_{\text{total}}(\mathbf{r}) = V_{\text{lattice}}(\mathbf{r}) + V_{\text{spin}}(\mathbf{r}) + \Delta V_{\text{Zeeman}}$. The phase shift in the exit wavefunction becomes

$$\phi(\mathbf{r}) = \frac{\sigma}{\hbar v}\int_0^t V_{\text{total}}(\mathbf{r}, z)\, dz + \frac{e}{\hbar}\int \mathbf{A} \cdot d\mathbf{r}$$

The exit wave function for magnetic crystal with spin wave order under magnetic field is thus: $\psi_{\text{exit}}(\mathbf{r}) = \psi_{\text{incident}}(\mathbf{r})e^{i\phi(\mathbf{r})}$

$$\psi_{\text{trans}}(\mathbf{r}) = \sum_{\mathbf{G}} A_{\mathbf{G}} e^{i(\mathbf{k}+\mathbf{G})\cdot\mathbf{r} + i\frac{e}{\hbar}\int \mathbf{A}\cdot d\mathbf{r}} + \sum_{\mathbf{q}} A_{\mathbf{q}} e^{i(\mathbf{k}+\mathbf{q})\cdot\mathbf{r} + i\frac{e}{\hbar}\int \mathbf{A}\cdot d\mathbf{r}} =$$

$$I(\boldsymbol{r}) = |\psi_{\text{trans}}(\mathbf{r})|^2 = \left|e^{i\frac{e}{\hbar}\int \mathbf{A}\cdot d\mathbf{r}}\right|^2 \left|\sum_{\mathbf{G}} A_{\mathbf{G}} e^{i(\mathbf{k}+\mathbf{G})\cdot\mathbf{r}} + \sum_{\mathbf{q}} A_{\mathbf{q}} e^{i(\mathbf{k}+\mathbf{q})\cdot\mathbf{r}}\right|^2$$

The Aharonov-Bohm phase contribution cancels out in the intensity expression shown in the final $I(\boldsymbol{r})$.

$$I(\boldsymbol{r}) = \left|\sum_{\mathbf{G}} A_{\mathbf{G}} e^{i(\mathbf{k}+\mathbf{G})\cdot\mathbf{r}} + \sum_{\mathbf{q}} A_{\mathbf{q}} e^{i(\mathbf{k}+\mathbf{q})\cdot\mathbf{r}}\right|^2$$

The Fresnel oscillations from crystal lattice symmetry remain unchanged by an external magnetic field, while the $\mathbf{q}$-vector in the second summation term depends on the field due to skyrmion or spin wave dispersion. This is the only field-sensitive component. Our experiments confirm that the external field modifies spin wave dispersion, that the observed LTEM contrast is purely a lattice effect. The field-induced changes in spin wave dispersion alter the peak positions, phases, and intensities of Fresnel oscillations, demonstrating that the observed LTEM contrast is magnetic in origin. Thus, Fresnel fringes originate from the interplay of lattice symmetry, skyrmion structures, and spin wave dispersion.

**Polarized Neutron Reflectometry (PNR) measurements:** were performed on the Magnetism Reflectometer at the Spallation Neutron Source at Oak Ridge National Laboratory[74-77]. Neutrons with wavelengths within a band of 2–8 Å and with a high polarization of 99% to 98.5% were used. Measurements were performed in a closed cycle refrigerator (Advanced Research System CCR) with an applied external magnetic field by using a 1.15 T Bruker electromagnet. Using the time-of-flight method, a collimated polychromatic beam of polarized neutrons with the wavelength band $\Delta\lambda$ impinged on the film at a grazing incidence angle, $\theta$, where it interacts with atomic nuclei and the spins of unpaired electrons. The reflections $R^+$ and $R^-$, were measured as a function of momentum transfer, $Q = 4\pi \sin\theta/\lambda$, for two neutron polarizations with the neutron spin parallel (+) or antiparallel (-) to the direction of the external field, $H_{ext}$. To separate nuclear from magnetic scattering, the data are presented as the spin-asymmetry ($SA$) ratio $SA$ = ($R^+$(Q) - $R^{-}$(Q))/($R^{+}$(Q) + $R^{-}$(Q)). A value of $SA = 0$ indicates that there is no magnetic moment in the system. Being electrically neutral, spin-polarized neutrons penetrate the entire multilayer structure and probe the magnetic and structural composition of the film through buried interfaces down to the substrate. PNR is a highly penetrating depth-sensitive technique to investigate the magnetic and chemical composition of the system with 0.5 nm resolution[78]. The NSLD and MSLD depth profiles correspond to the depth distribution of the chemical and IP magnetization vector distributions at the atomic scale, respectively. Spin-polarized neutrons at grazing incidence penetrate the entire multilayer film down to the substrate and provide depth-resolved information about the structure and the magnetization vector of the film through buried interfaces. Note that neutron interactions with Eu atoms are affected by unpaired *f*-electrons with unique magnetic properties. The distinct characteristics of Eu lead to variations in the neutron absorption scattering length density (ASLD), in contrast to other elements. It is essential to closely examine the distinctive properties of materials containing Eu atoms while analysing their neutron absorption behaviour[59]. To spatially resolve the net magnetization vector through the depth of trilayer configurations, we used a time-of-flight method using a highly collimated polychromatic beam of polarized neutrons impinging on the film under grazing incidence and interacting with atomic nuclei and the spins of unpaired electrons providing the information on the depth profile of the nuclear and magnetic scattering length densities (NSLD and MSLD) corresponding to chemical and in-plane magnetization vector distributions, respectively[74-77].

**Micromagnetic simulations:** we propose a model of a hopfion ring encircling the skyrmion lattice based on the trilayer system (EuS-$Bi_2Se_3$-EuS) shown in Fig. 1. Using MuMax3 software[79] based on LLG function, we conduct micromagnetic modelling to investigate the feasibility of constructing such a system using the following material parameters: saturation magnetization $M_{sat}$ = 37.6×10$^{4}$ A m$^{-1}$, Heisenberg exchange constant $A_{ex}$ = 1.94×10$^{-14}$ J m$^{-1}$, interfacial constant $DMI$ = 4.56×10$^{-6}$ J m$^{-2}$ and Gilbert damping constant $\alpha$ = 0.3. The mesh is set to 1×1×1 nm$^{3}$ for a 100, 300, and 500 nm$^{2}$ geometry, with each EuS and $Bi_2Se_3$ layers having a thickness of 5 nm. For each top and bottom interface layer of EuS, $M_{sat}$ was reduced to one-third of its original value and the simulation was relaxed for the hopfion-skyrmion system.

Having a narrower hopfion ring would not survive against the edge effects and dipolar fields from the skyrmion lattice. The anisotropy field is chosen as $K_{u1}$ = 1.0×10$^{6}$ J m$^{-3}$ in the out-of-plane (+**z**) direction. Due to this high value, we observe little or no relaxation of the magnetic features as this anisotropy field overcomes the demagnetization field. Hopfions are expected to show chirality in three dimensions, and the moments surrounding the ring will swirl around it[80]; however, due to the high uniaxial anisotropy field, we do not observe these effects (Extended Data Figs. 7 and 8). The width of the hopfion keeps it intact even in proximity to the skyrmion lattice, and despite the strong demagnetization or dipolar interactions between them. If the magnetic features were allowed to relax further, one could expect the hopfion to lose its symmetric shape and the shape of its domain walls extending to the edge of the geometry due to edge interactions (Extended Data Fig. 8). The modelling identified the material parameter ranges ($M_{sat}$ and *DMI*) needed for stabilizing the skyrmion lattice with and without the hopfion ring. These ranges are presented in the Extended Data Figs. 4 and 5 for irregular hexagon islands.

**X-ray magnetic circular dichroism (XMCD):** The samples were examined using XMCD to verify interfacial magnetic coupling. At the Advanced Photon Source's beamline 4-ID-C, we conducted a number of soft X-ray absorption spectroscopy investigations by simultaneously measuring the bulk-sensitive fluorescence yield and the surface-sensitive total electron yield. Initially, we investigated the magnetic condition of the EuS layer. With a valence state of 2+, the EuS layer exhibits a significant XMCD, and its lineshape is in line with a local moment of 7 $\mu_B$ per atom. Extended Data Fig 10 displays the XMCD spectra for the $Eu^{2+}$ state as a function of photon energy as a function of temperature to obtain the magnetic properties of Eu atoms after the moments are aligned by an applied perpendicular 5 T field training. The scans shown were averaged over many scans and analysed with X-ray energy over the Eu $M_5$-edge (~1128 eV, $3d_{5/2} \rightarrow 4f$ transition) at a step resolution of ~0.2 eV. The magnetic information derived from the XMCD is related to the 7 $\mu_B/Eu^{2+}$ ion moment.

Every experiment was thoroughly examined, with conclusions drawn from supporting data from further experiments. A definitive, trustworthy result depends critically on the quality of the samples. It is evident from our experimental results, using various approaches and high-quality materials that interfacial magnetism exists on both the short- and long-range scales up to ambient temperature[58,81].

**EXTENDED DATA FIGURES**

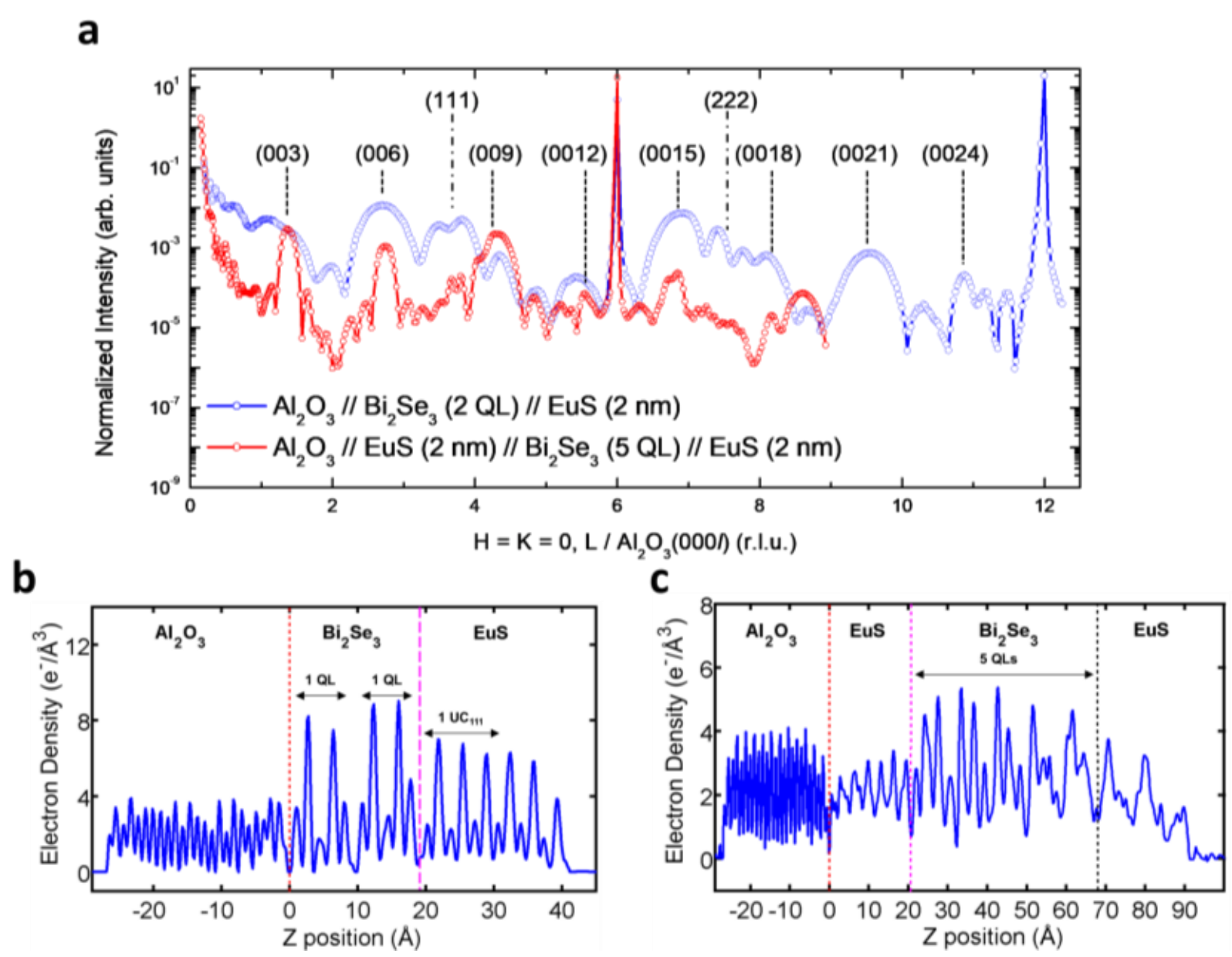

**Extended Data Fig. 1 | X-ray diffraction analysis for epitaxial bi- and tri-layer films grown on sapphire**. **a**, In order to shed light on the interface microstructure we performed sets of crystal-truncation rod (CTR) measurements and employed coherent Bragg rod analysis (COBRA) to determine the real space electron density profile across the interfacial region with atomic precision[72]. The CTR measurements were performed with a six-circle diffractometer. The specular CTR (00*L*) of the bilayer and trilayer samples configuration were analysed by the COBRA method. From the density profile, **b**, for bilayer and **c**, trilayer, the interface between EuS and $Bi_2Se_3$ layers are well defined either in bilayer or trilayer. Sharp interface transition is shown without any chemical inter-diffusion by resolving the interface bond, clearly. Se atoms are in direct contact with the last oxygen layer of the sapphire. The $Bi_2Se_3$ is terminated with the Se layer at the interface and the S layer is in direct stacking with the Se layer on top of it via interfacial bonding separation by ~2.45 Å. Overall, EuS (111) on $Bi_2Se_3$–Sapphire shows a good epitaxial cube-on-hexagon growth relation.

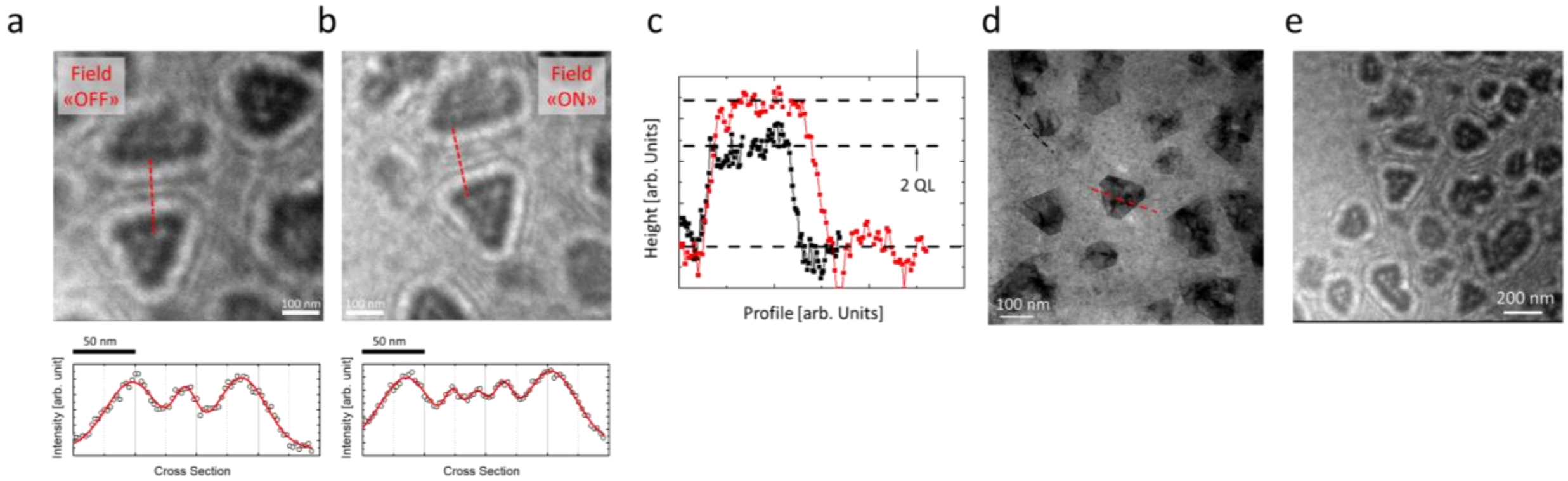

**Extended Data Fig. 2 a,** Hopfion and spin density wave formation. There is a discernible periodicity in the excitation between two separate island spin density waves when the field is off. **b**, A perpendicular field causes both the hopfion diameter and the wave periodicity to decrease. It also shows the potential pattern of interference. The line cuts are shown below for each. For the "Field-ON" configuration, the applied out-of-plane component of the magnetic field is 1.53 T. **c**, Height profile obtained from the line cut indicated by red and black dashed lines in **d**, revealing the characteristic truncated island morphology of epitaxially grown $Bi_2Se_3$ with thickness variations of ~2 QL between adjacent islands. **d**, Top-view TEM image showing the surface morphology of the trilayer layer film, with visible island formations characteristic of van der Waals material growth. **e**, LTEM image, displaying magnetic contrast that remains consistent across islands of varying thickness, demonstrating that the observed magnetic phenomena are intrinsic to the EuS / $Bi_2Se_3$ / EuS heterostructure.

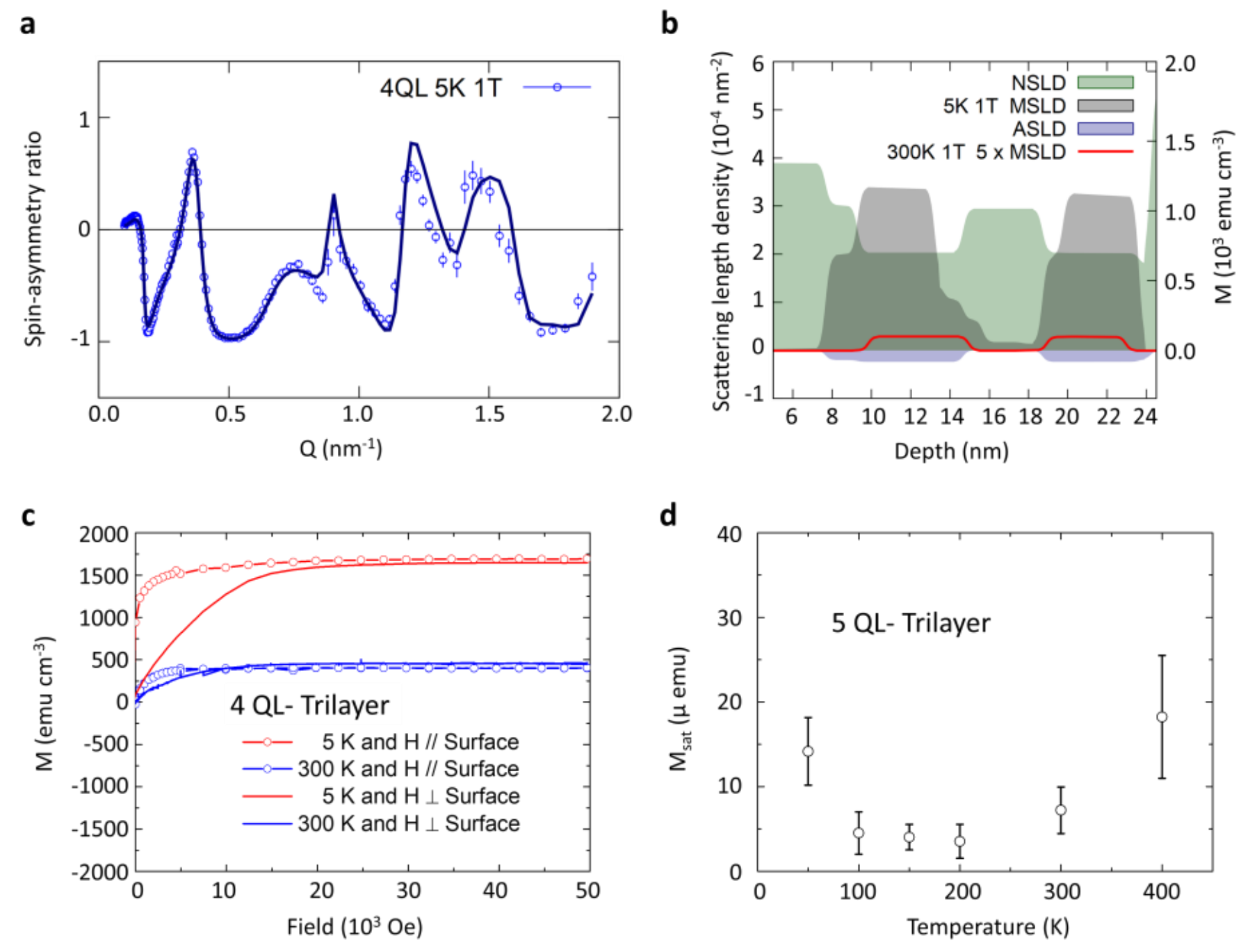

**Extended Data Fig. 3 | PNR and SQUID magnetometry measurements for trilayer EuS–$Bi_2Se_3$–EuS,** PNR results for 4 QL-trilayer (EuS (5 nm) – $Bi_2Se_3$ (4 QL) – EuS (5 nm)) at 5 K in comparison with 300 K data. In **a**, the corresponding spin-asymmetry (SA) ratio and model fits are displayed with solid lines, SA = $(R^+ - R^-)/(R^+ + R^-)$, is derived from the reflectivity measurement fitting for 4 QL-trilayer sample. In **b**, neutron nuclear (NSLD, green), magnetic (MSLD, grey) and absorption (ASLD, purple) scattering length density profiles are shown along the trilayer portion of the epitaxial sample which were recorded at 5 and 300 K with an in-plane 1 T field. In **c**, measurements of M(H) at 5 and 300 K temperature in a parallel and perpendicular field configurations for 4 QL-trilayer (EuS (5 nm) – $Bi_2Se_3$ (4 QL) – EuS (5 nm)) sample. Compared to the sample shown in Fig. 3, this one is distinct. This has a couple orders of magnitude larger saturation moment at 300 K, however the low temperature values is 25% lower. In **d**, the saturation moment versus temperature data is shown for 5 QL-trilayer sample.

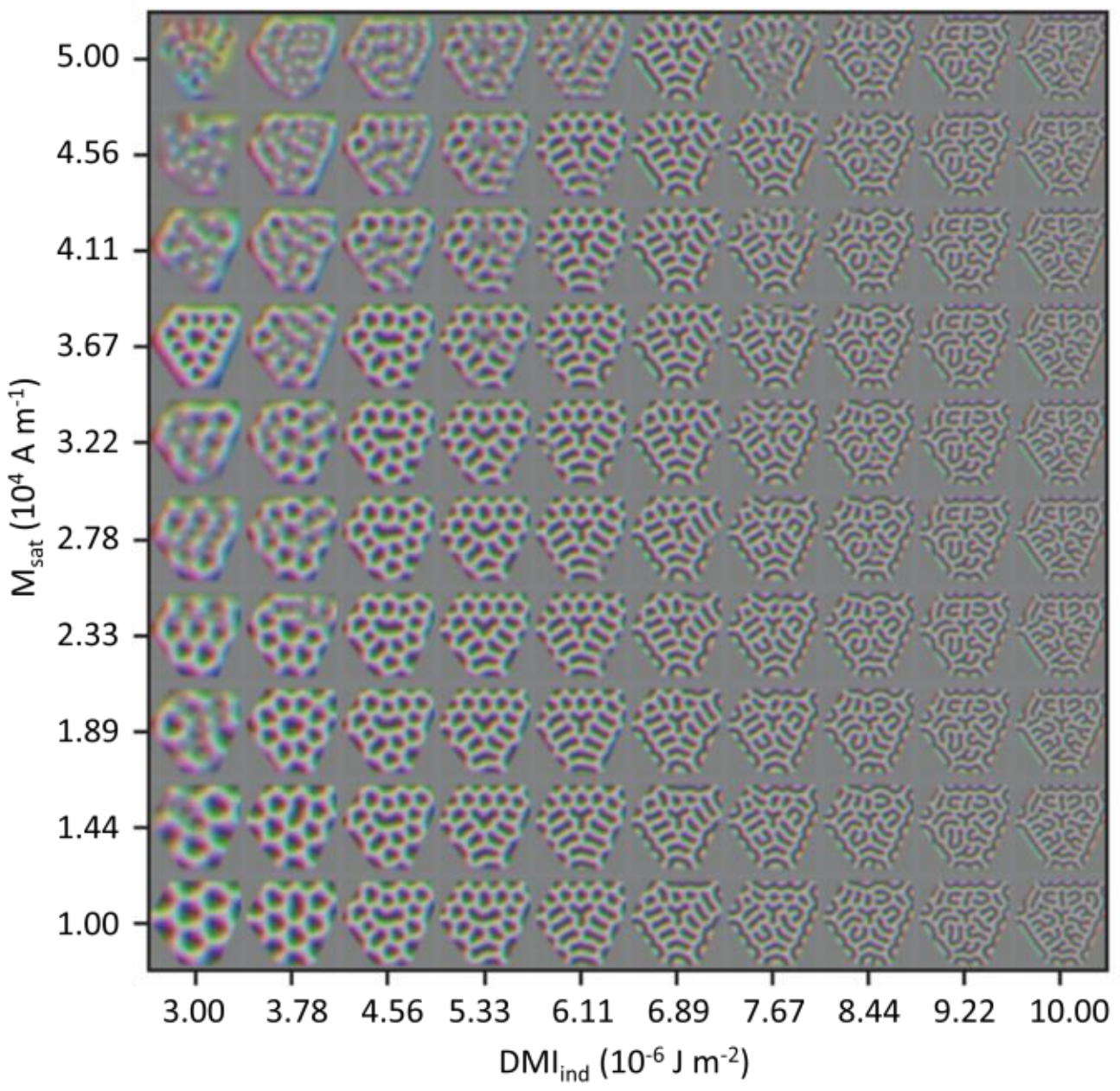

**Extended Data Fig. 4 | Micro Magnetic Modelling without hopfion**, the effect of change in the DMI and saturation magnetization on the stabilization of a single skyrmion island without hopfion ring.

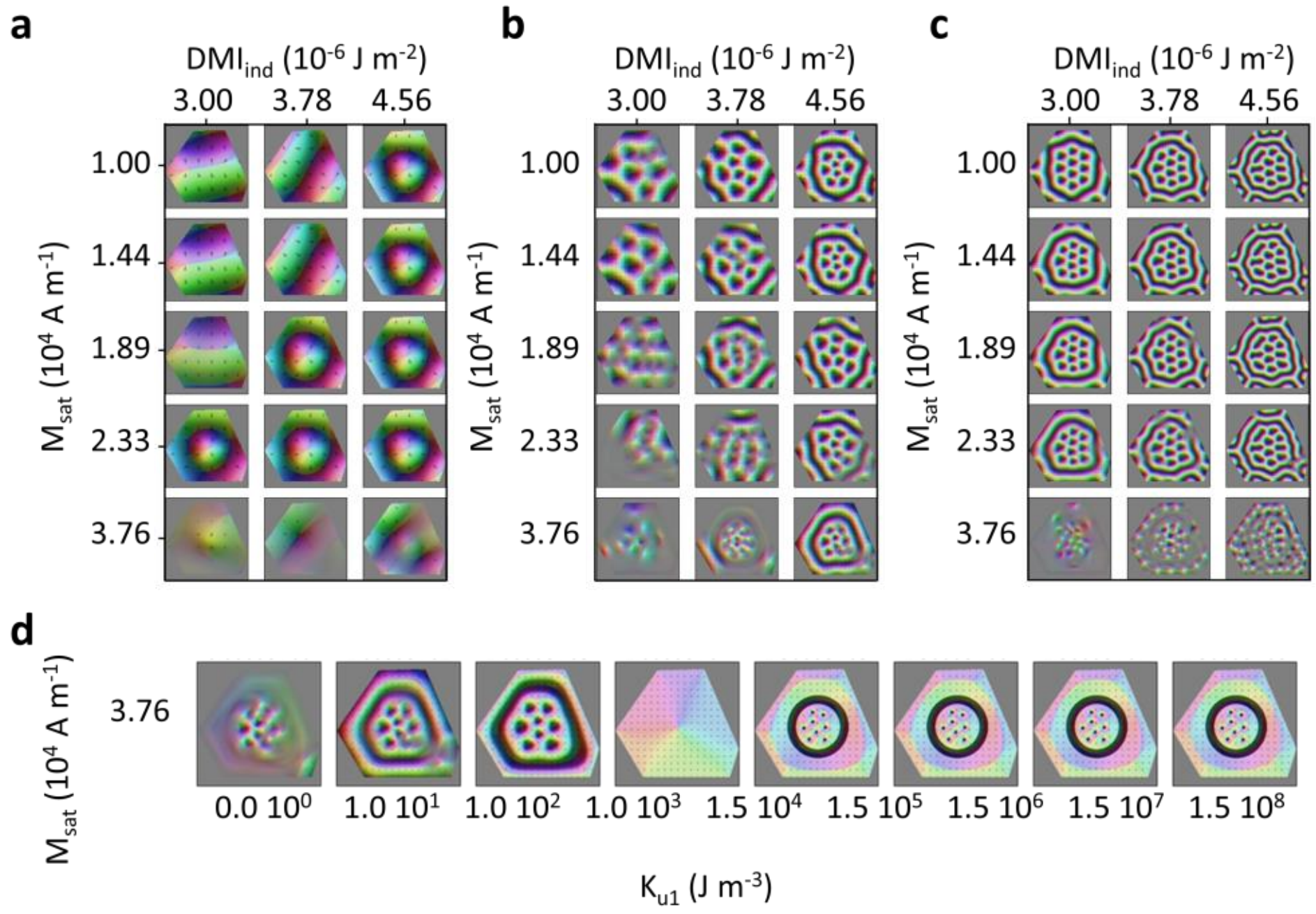

**Extended Data Fig. 5 | Micromagnetic simulations of magnetic texture with hopfion rings for trilayer films.** The corresponding simulations for Fig. 2a were run on irregular hexagon geometry with sizes at 100 in **a**, 300 in **b**, and 500 nm$^2$ in **c**. The entire isosurfaces of the skyrmion and hopfion lattice construction are matched with experimental data in **b**, due to their accurate size and shape by convenient $M_{sat}$ and *DMI*

parameters. In **d**, a large range was also run for the corresponding uniaxial anisotropy parameters $K_{u1}$ with $DMI_{ind}$ = 4.56 × $10^{-6}$ J $m^{-2}$, $A_{ex}$ = 1.94 × $10^{-14}$ J $m^{-1}$, and $M_{sat}$ = 37.6 kA $m^{-1}$ with regular triangle geometry with sizes at 300 $nm^2$.

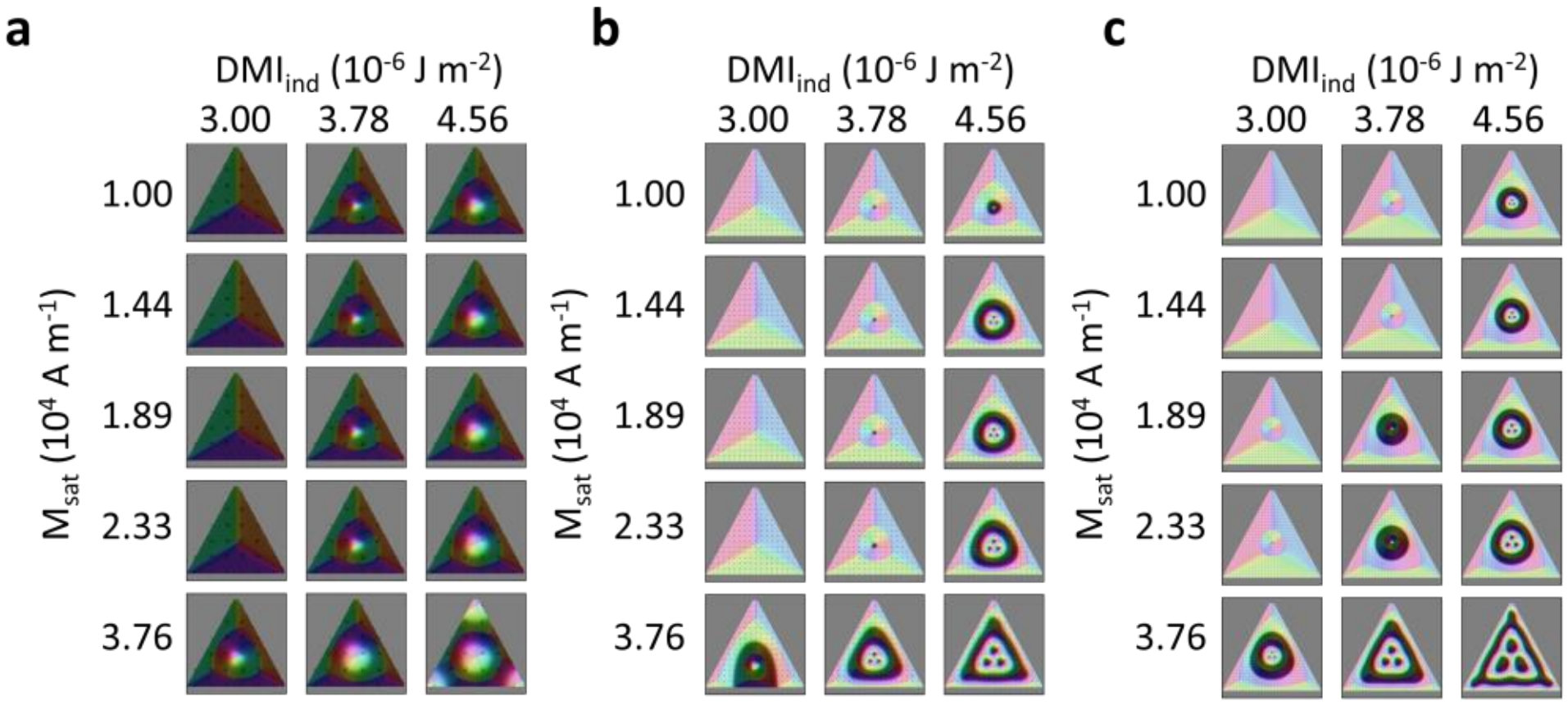

**Extended Data Fig. 6 | Micromagnetic simulations of 3 skyrmions with hopfion ring formation for trilayer films.** The corresponding simulations for Fig. 2g were run on regular triangle geometry, which includes 3 skyrmions, with sizes at 100 in **a**, 300 in **b**, and 500 $nm^2$ in **c**. The entire isosurfaces of the skyrmion and hopfion lattice construction are matched with experimental data in **b**, due to their accurate size and shape by convenient $M_{sat}$ and *DMI* parameters. A large range was also run for the corresponding uniaxial parameters with *DMI* = 4.56 × $10^{-6}$ J $m^{-2}$, $A_{ex}$ = 1.94 × $10^{-14}$ J $m^{-1}$, and $M_{sat}$ = 37.6 kA $m^{-1}$.

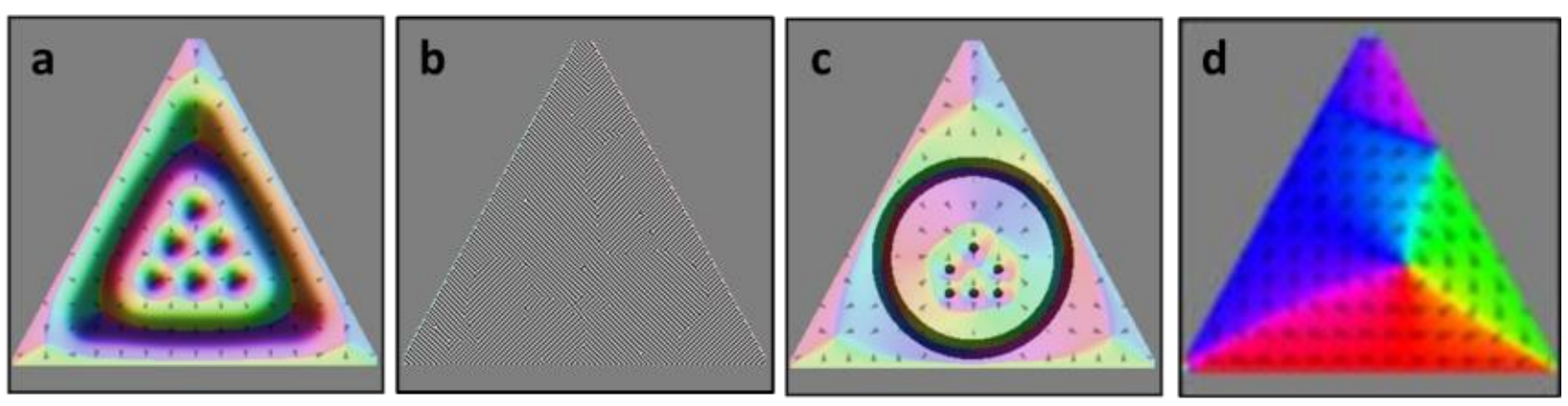

**Extended Data Fig. 7 | Micromagnetic model on extended parameters for key material parameters.** The effect of increasing a few orders of magnitude of the key material parameters in our study from those in the initial model shown in **a**, where we obtain *DMI* = 4.56 µJ $m^{-2}$, $M_{sat}$ = 37.6 kA $m^{-1}$, and $K_{u1}$ = 1×$10^3$ ($A_{ex}$ is fixed in all figures at 1.94×$10^{-14}$ J $m^{-1}$). Increasing the interfacial *DMI* strength to 1 mJ $m^{-2}$ converts the skyrmion lattice and hopfion ring into a magnetization state of a strip-like domain as we allow them to relax in the micromagnetic model as shown in **b**. In **c**, having a high anisotropy constant, $10^5$ J $m^{-3}$, a couple of orders of magnitude larger prevents the initial magnetization state from relaxing, and we observe the high symmetry shapes of the skyrmion lattice and hopfion ring. The strong perpendicular uniaxial anisotropy field inhibits the moments from canting or reorienting themselves and overpowers the shape

anisotropy effects. Having an order of magnitude larger $M_{sat}$, 376 kA m$^{-1}$, in **d**, increases the demagnetization field effect (allows for large domain formation) and shape anisotropy, dissolving the magnetic features in the ferromagnetic layers.

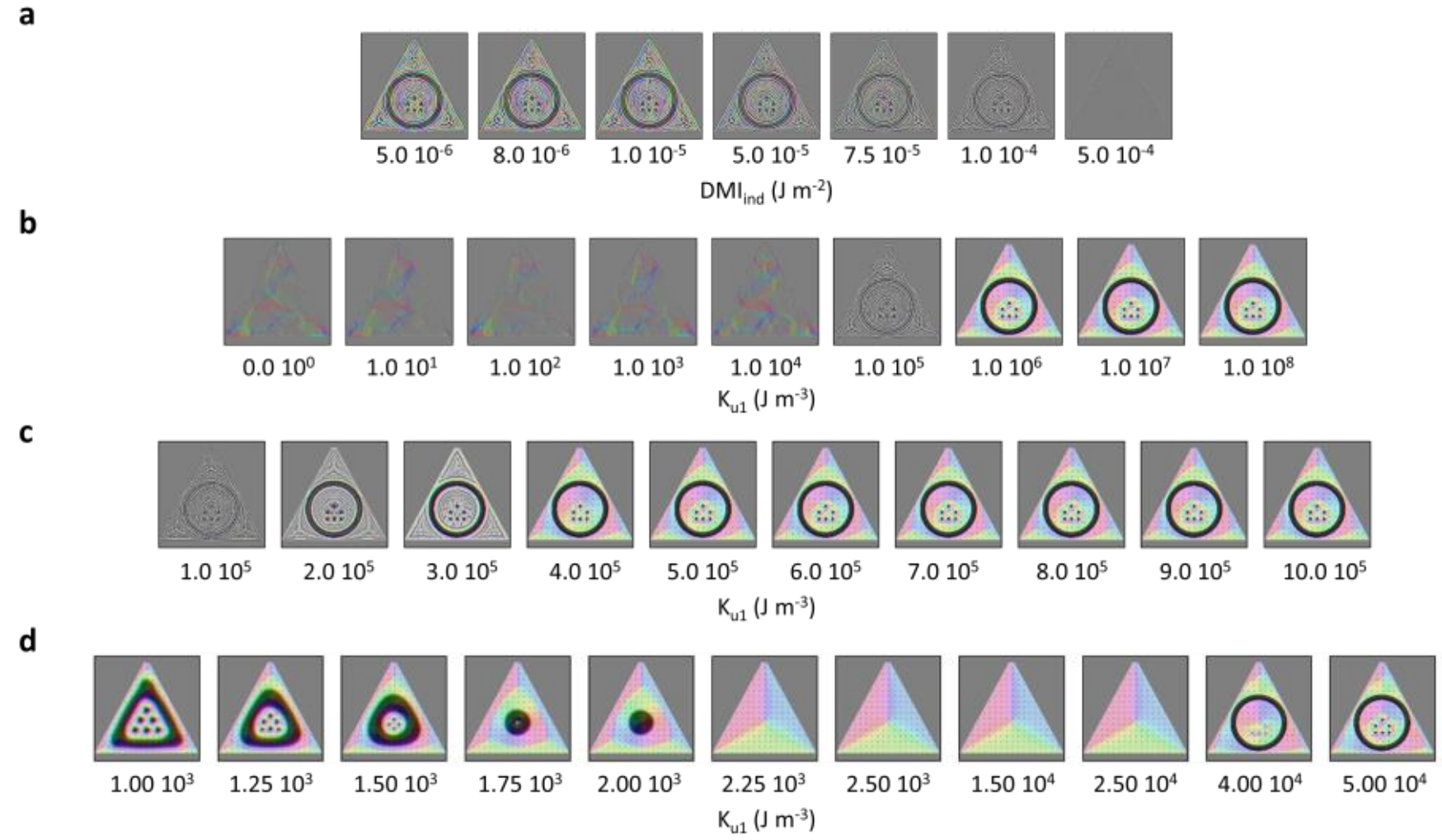

**Extended Data Fig. 8 | Micromagnetic model on extended parameters.** In **a,** the effect of introducing the *DMI* constant between 5.0 μJ m$^{-2}$ to 50 mJ m$^{-2}$ for the magnetic features with high saturation magnetization $M_{sat}$ = 800 kA m$^{-1}$ and high uniaxial anisotropy $K_{u1}$ = 1×10$^{5}$ J m$^{-3}$ are shown. High $M_{sat}$ increases the magnetostatic interactions and the demagnetization energy that induces strip-like domain formation. Increasing the *DMI* allows the strip-like domain system to dominate around these features and eventually converts the ring and skyrmion lattice into a system of such domains. In **b**, results show that keeping $M_{sat}$ and *DMI* high at 800 kA m$^{-1}$ and 10 mJ m$^{-2}$, respectively, while increasing the uniaxial anisotropy constant restores the ring and skyrmion lattice initial magnetization states, and the strip-like domain system disappears, which also shown for a finer sweep of $K_{u1}$ in **c**. In **d**, finer sweep for the region at the vicinity of $K_{u1}$=1.5 x 10$^{4}$ J m$^{-3}$ with *DMI* = 4.56 × 10$^{-6}$ J m$^{-2}$, $A_{ex}$ = 1.94 × 10$^{-14}$ J m$^{-1}$, and $M_{sat}$ = 37.6 kA m$^{-1}$ with regular triangle geometry with sizes at 300 nm$^{2}$.

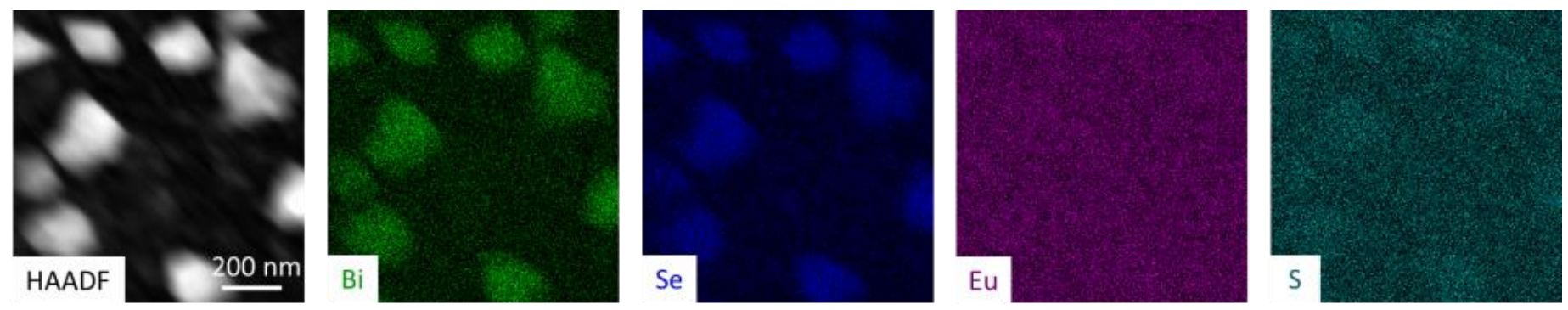

**Extended Data Fig. 9 | HAADF and EDS measurements for trilayer samples grown on a grid.** Energy dispersive spectroscopy (EDS) map displaying the spatial atomic distribution of a trilayer sample grown on $Si_3N_4$ membrane. The elements Bi, Se, Eu, and S are evenly dispersed throughout the film surface, based on the element mapping pictures. The element molar ratios of Eu:S and Bi:Se are around 1:1 and 2:3, respectively, as anticipated for grown materials. Because of their lower packing ratio, crystallite areas contrast less than polycrystalline regions. The presence of distinct domain sizes is validated using high-angle annular dark field (HAADF) pictures, in which the contrast of the images is approximately correlated with the square of the atomic number of the chemical species under investigation. The temperature of deposition affects both the domain size distribution and their inter-domain spaces. At low temperatures, the domain sizes decrease down to 2–10 nm; nevertheless, at high temperatures, the situation drastically shifts and causes a notable rise up to a few of hundred nm. The sharp rise in crystalline sizes suggests that either larger domains are directly growing, or tiny domains are coalescing because of the species' greater mobility at higher temperatures.

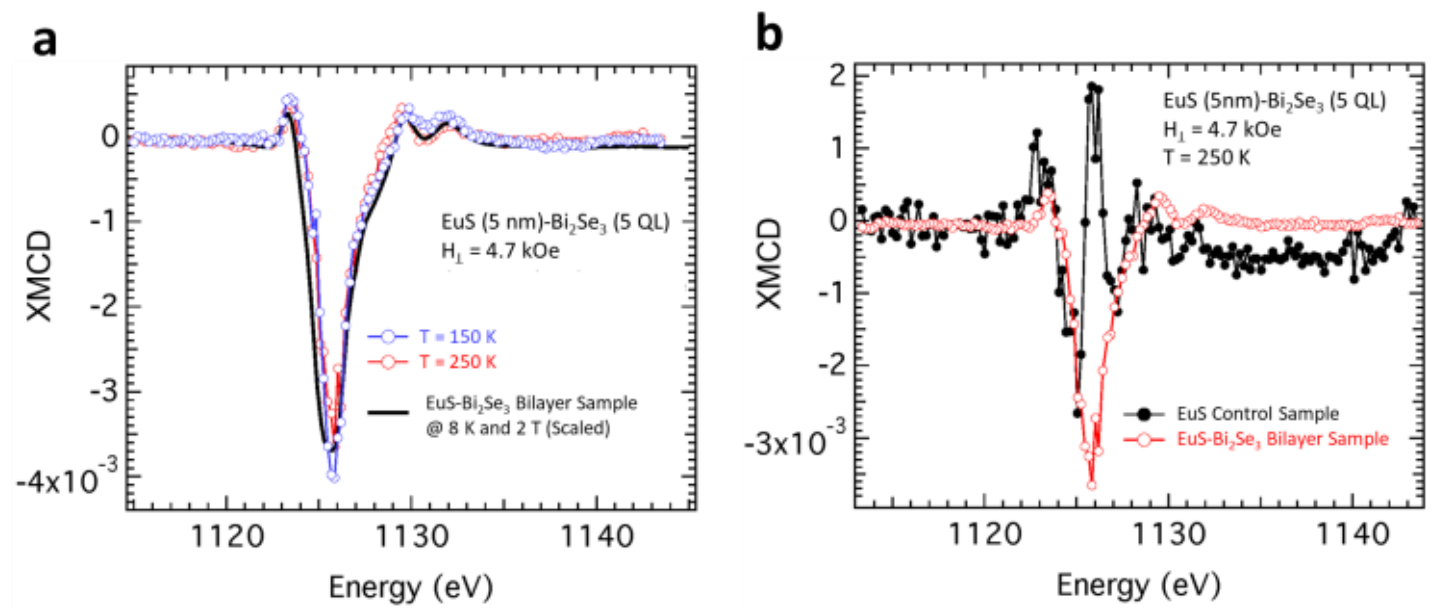

**Extended Data Fig. 10 | High-Temperature XMCD data analysis for EuS-$Bi_2Se_3$ interface.** The XMCD (X-ray Magnetic Circular Dichroism) spectra as a function of photon energy for $Eu^{2+}$ state is shown, where the spectra are taken as a function of temperature to get magnetic features of Eu atoms after aligning the moments by 5 T applied field for both EuS (5 nm) – Sapphire in **b** and EuS (5 nm) – $Bi_2Se_3$ (5 QL) – Sapphire samples in **a** and **b**. The depicted scans were evaluated with X-ray energy over Eu $M_5$-edge (~1128 eV, $3d_{5/2} \rightarrow 4f$ transition) using ~0.2 eV step resolution and averaged out from several scans. From the

temperature dependence measurements, according to Curie-Weiss at 250 K, it is expected to have 60% of the 150 K value, however, at 250 K, we observed 83% of the 150 K value where similar behaviour was observed with SQUID magnetometry measurements[8]. Given that the control sample's signal in **b** deviates from Eu's line shape, it is most likely a derivative artifact from sample charging.

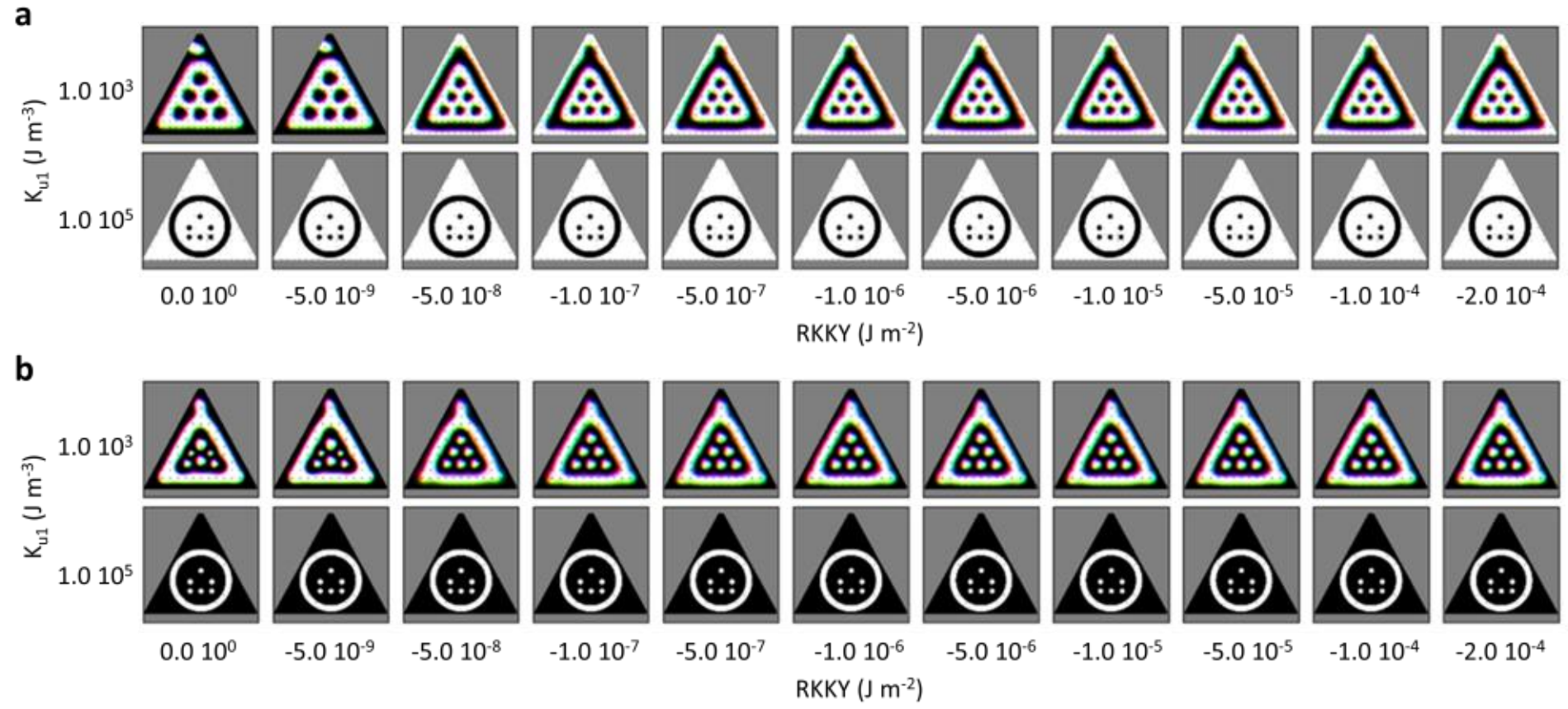

**Extended Data Fig. 11 | Micromagnetic model for interlayer exchange coupling and RKKY for trilayer system**. In **a**, the bottom layer and in **b** the top layer z-slice image of the skyrmion lattice simulated results are shown. Here, we introduced the antiferromagnetic coupling (AFM) between the top and bottom ferromagnetic layers of EuS. The Ruderman-Kittel-Kasuya-Yosida (RKKY) constant was swept for the range shown Figs. **a**, and **b** for two values of out-of-plane uniaxial anisotropy constant $K_{u1}$. The effect was mediated through the scaled exchange constant between the top and bottom interfaces that were initialized with a Neel skyrmion lattice and a hopfion ring with z-plane mirrored magnetization profiles. The rest of the material parameters used in these runs were as before. The RKKY constant was varied from 0 to -0.2 mJ $m^{-2}$, where the strength of AFM coupling can be a function of the spacer layer thickness[82] and can act proportional to the TI layer thickness[63].

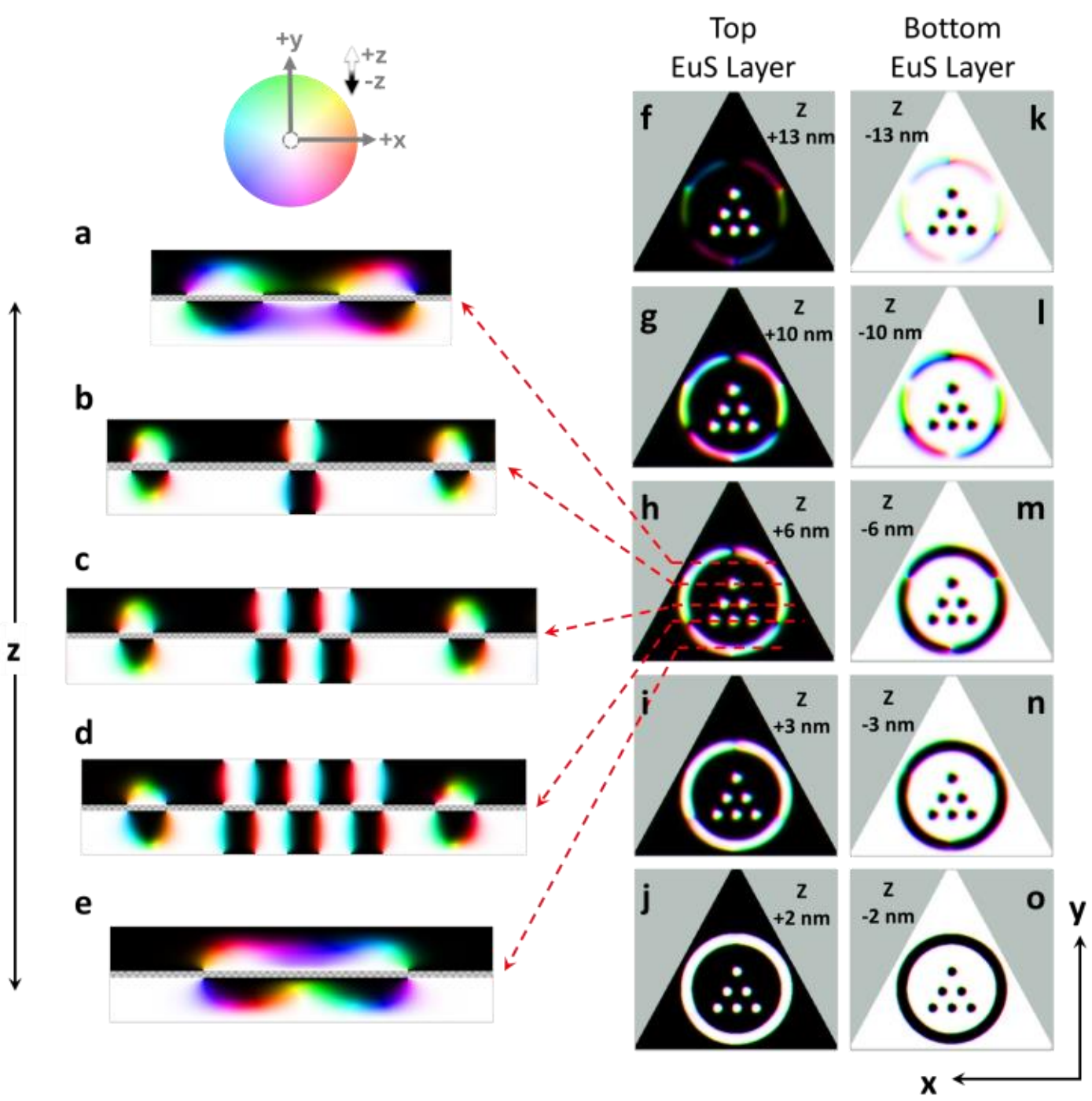

**Extended Data Fig. 12 | Extended micromagnetic simulation across the entire thicker trilayer heterostructure**. The micromagnetic simulation of EuS (20 nm) / $Bi_2Se_3$ (5 QL) / EuS (20 nm) shows a 3D representation across the thickness of each EuS layer. The simulation illustrates the skyrmion lattice surrounded by the hopfion in the antiferromagnetically coupled EuS layers through $Bi_2Se_3$ layer, which is not explicitly included in the simulations and is marked as a grey-crossed area. The phase profile along x-z view at different y positions is shown in **a** to **e,** while the lateral view is depicted in the x-y plane in **f** to **o**. Lateral cross-sections were taken at different EuS thicknesses, specifically at ±2, ±3, ±6, ±10, and ±13 nm, corresponding to their distances from each $Bi_2Se_3$ interface, where + and - signs indicate for the top and bottom EuS layers, respectively. The color bar shows the spin phase changes for increasing azimuthal and polar angles.